\documentclass{aastex62}
\usepackage{amsmath}
\graphicspath{{./}{figures/}}

\received{2019 June 6}
\accepted{2019 August 25}
\submitjournal{AJ}

\shorttitle{Planetary reflected light with ICA}
\shortauthors{Di Marcantonio et al.}

\begin{document}

\title{Using Independent Component Analysis to detect exoplanet reflection spectrum from composite spectra of exoplanetary binary systems}

\correspondingauthor{Paolo Di Marcantonio}
\email{paolo.dimarcantonio@inaf.it}

\author[0000-0003-3168-2289]{Paolo Di Marcantonio}
\affiliation{INAF - Osservatorio Astronomico di Trieste, Via G. B. Tiepolo 11, Trieste, I--34143, Italy}

\author[0000-0002-3319-6375]{Carlo Morossi}
\affiliation{INAF - Osservatorio Astronomico di Trieste, Via G. B. Tiepolo 11, Trieste, I--34143, Italy}

\author[0000-0001-5611-2333]{Mariagrazia Franchini}
\affiliation{INAF - Osservatorio Astronomico di Trieste, Via G. B. Tiepolo 11, Trieste, I--34143, Italy}

\author[0000-0003-3449-4355]{Holger Lehmann}
\affiliation{Th\"{u}ringer Landessternwarte Tautenburg, Sternwarte 5, 07778 Tautenburg, Germany}




\begin{abstract}
The  analysis of the wavelength-dependent albedo of  exoplanets
represents a direct way to provide insight of their atmospheric composition
and  to constrain theoretical planetary atmosphere modelling.
Wavelength-dependent albedo can be inferred from the exoplanet's reflected light of the 
host star, but this is not a trivial task In fact, the planetary signal  may be several orders of magnitude lower ($10^{-4}$ or below)
than the flux of 
the host star, thus making its extraction  very challenging.  Successful
detection of  the planetary signature of 51~Peg\,b has been recently obtained  by using cross-correlation function (CCF)
or autocorrelation function (ACF) techniques.
In this paper we present an alternative
method based on the use of Independent Component Analysis (ICA). 
In comparison to the above-mentioned techniques, the main advantages of ICA are that the extraction is
\textit{``blind''} i.e. it does not require any \textit{a priori} knowledge of the underlying signals, 
and that our method allows us not only to detect the planet signal  but also to estimate its wavelength dependence.
To show and quantify the effectiveness of  our method we successfully applied it to  both simulated data and  real data of an eclipsing  binary star system.
Eventually, when applied to real 51~Peg~+~51~Peg\,b data, our method extracts the signal of 51~Peg but we could not soundly detect the reflected spectrum of 51~Peg\,b mainly due to the insufficient $SNR$ of the input composite spectra. Nevertheless, our results show that with ``ad-hoc'' scheduled  observations an ICA approach will be, in perspective, a very valid tool for studying exoplanetary atmospheres. 
\end{abstract}

\keywords{methods: data analysis -- methods: statistical  -- planets and satellites: atmospheres -- binaries: eclipsing -- techniques: radial velocities -- techniques: spectroscopic}


\section{Introduction} 
\label{sec:intro}

  Observations of  spectra of stars with exoplanets are a fundamental tool to determine not
only the basic orbital parameters of  surrounding planets but also to get hints on their atmosphere (if any). In fact,  spectroscopic observations contain  the variations of Doppler shifts 
of spectral lines of both  components, i.e. star and planet(s), caused by changes of their radial velocities during   their  orbital motion \citep[for discussion about radial velocity see][]{Lindegren03}.
Furthermore, the radius and the  mass of exoplanets can be derived by combining the  light-curve signal of  exoplanetary transits with radial velocities using Kepler laws.

To derive information on  exoplanet albedo and spectrum, we need 
to distinguish, in the observed (composite) spectra, the spectral features belonging to 
individual system components, i.e. to decompose the observed superposition of their lights.
Since  information on the individual spectra of the observed system is entangled in
the composite spectra, the procedure of decomposing   them is often referred  to as  {\it disentangling}.
 {\it Disentangling} can be performed by comparing spectra taken at different known radial velocities of the star and of the exoplanet or at phases with different known light ratios 
(e.g. during transit and/or eclipse). These approaches are also at the basis of the studies of
binaries \citep[see for example the discussion by][and references therein]{Hadrava16}.

In  case of exoplanet the reflected spectral signal from its atmosphere, which should mimic the stellar signal, permits to gather the planetary albedo and thus info on its atmospheric composition.
 In fact, the planetary reflectivity depends on wavelength because of scattering and absorption processes that suffer the incoming radiation  in its atmosphere and/or surface. The subject of modelling and understanding the planetary atmospheric structure from the reflected and emitted spectra is very complex, and there are many different approaches (see an overview of the problem in \citet{Marley15} and a more detailed review in \citet{Marley13}.
Moreover, the determination of the planetary albedo is not a simple task from the observational point of view because of the extremely low flux ratio between the reflected light from the planet and the stellar one. As examples of the several efforts  to derive exoplanetary albedos we recall \citet{Charbonneau99,CollierCameron99,Rodler10}. In most of the cases the authors were able to establish only upper limits for the reflected signals. Recently a significant improvement has been obtained by means of new ad-hoc techniques.
\citet{Martins13} proposed a technique that makes use of the Cross Correlation Function (CCF) of high resolution spectra to amplify the  planetary signal above  stellar noise and detected the reflected signal of 51\,Peg\,b. 
Moreover, \citet{Martins18} showed, by using simulated data, that CCF method
can successfully recover the geometric albedo of exoplanets over a wavelength range.
\citet{Borra18} suggested the use of  autocorrelation function (ACF) as a valid alternative to  CCF, in particular because it does not require a weighted binary mask. Both techniques search for secondary maxima in  the CCF or ACF of the composite spectra due to the presence of a Doppler shifted reflected light signal.

In this paper we propose an alternative technique to both CCF and ACF that uses the Independent Component Analysis \citep[ICA;][]{Hyvarinen01} as an effective
way to {\it disentangle} the exoplanetary signal from  instrument
and stellar  components.
ICA is a method for extracting hidden components from multivariate statistical data that are both  statistical independent and nongaussian (see Section\,\ref{sec:ica}).
The use of  ICA in this paper is based on the assumption that the individual spectra forming the composite spectrum of the stellar-planet system  can be considered as two independent components (see for discussion  Section\,\ref{sec:simul}).
Since ICA  was initially proposed to solve the blind source separation (BSS) problem, one of the main characteristics of our method, hereafter Exoplanet Reflected Light via Independent Component Analysis (ERLICA), lies in the fact that it does not require any \textit{a priori} auxiliary information  but only the observed data
themselves.
 Furthermore, our method not only enable us to detect a planetary signal but is also able to provide, if the signal to noise ratio (SNR) of the input spectra is enough, the variation of the planet albedo with wavelength. 

The goal of this work is to assess and analyze the efficiency of our method in
detecting a planetary signal and in estimating its wavelength dependency  by means of  reflected starlight. 
To do so, we apply ERLICA to  both artificial (simulated) and real (observed) spectra in the same optical region studied by \citet{Martins15} and by \citet{Borra18} in order to be able to compare our and their results. Obviously, the same kind of analysis could have been applied to other wavelength region, e.g. to IR or UV data.

In Section~\ref{sec:method} we summarize the theoretical framework of an exoplanetary system and give a short description of the ICA packages (FastICA\footnote{https://research.ics.aalto.fi/ica/fastica/} and ICASSO\footnote{https://research.ics.aalto.fi/ica/icasso/}) we use. Section~\ref{sec:simul} describes the adopted method and its validation by using  simulated  data for the 51\,Peg\,$+$51\,Peg\,b system.
Section\,\ref{sec:planet} shows and discuss the results of our methodology applied to the real (observed) case of the binary eclipsing star  R\,CMa (an Algol system) and of 51~Peg~$+$~51~Peg\,b system.
Conclusions are given in Section~\ref{sec:conclusion}.

\section{Theoretical framework} \label{sec:method}

\subsection{Exoplanet's reflected light} \label{sec:phase}

The model adopted in this paper to simulate the photometric variations of
the stellar  light reflected from a planet  orbiting its host star is based on the discussion 
presented in \cite{Charbonneau99} and reused by \cite{Martins13}. 
The phase-dependent flux ratio of the planet flux
$F_{\rm planet}$ relatively to the star flux $F_{\rm star}$ is given by

\begin{equation} \label{eq:fluxratio}
    \frac{F_{\rm planet}(\alpha)}{F_{\rm star}} = p g(\alpha) \left(\frac{R_{\rm p}}{a}\right)^2
\end{equation}
where $p$ is the geometric albedo of the planet, $R_{\rm p}$ is the planet radius,
$a$ is the planet orbital distance and $g(\alpha)$ is the phase function. It holds for the case  $R_{\rm p} \ll R_{\rm star} \ll a$. As described in 
\cite{Borra18}, the phase angle, $\alpha$, is the angle between the star 
             and the Earth when seen from the planet. It depends on the 
	     position of the planet in its orbit and takes value in range
	     $0\le\alpha\le 180^\circ$.  For an orbital inclination $i$ and orbital phase $\phi$, assuming a circular orbit, $\alpha$ can be derived from

\begin{equation} \label{eq:phi}
    \cos(\alpha) = - \sin(i) \cos(2\pi\phi)
\end{equation}
The orbital phase $\phi$ in equation~\ref{eq:phi} traces the position of the planet on its orbit around the star. $\phi$ takes values between 0 and 1. In \cite{Charbonneau99} it is assumed $0$ at the time of maximum radial velocity, in other works (e.g.~\cite{Borra18}), phase $0$ corresponds to the time of the mid-point of the transit. In this paper we will use the latter definition, i.e.
for us $\phi=0$ at the  mid-point of the transit or of the inferior conjunction. 
For this particular choice, at a given time $t$, $\phi$ value can be calculated knowing the planet time of transit $t_{\rm 0}$ and its orbital period $P_{\rm orb}$ by

\begin{equation} \label{eq:fi}
    \phi = \frac{t-t_{\rm 0}}{P_{\rm orb}}
\end{equation}
The phase function $g(\alpha)$ in equation~\ref{eq:fluxratio} models the fraction of the maximum planet's reflected flux at full phase. It varies, depending on the planet position on its orbit, since we see different portion of the planet illuminated hemisphere. As described in \cite{Langford11}, in the simplest case, we can assume an isotropic scattering over $2\pi$~sr (Lambert-law sphere) in which case $g(\alpha)$ has typically the form

\begin{equation} \label{eq:galfa}
    g(\alpha) = \frac{[\sin(\alpha) + (\pi - \alpha) \cos(\alpha)]}{\pi}
\end{equation}
and $\alpha$ is related to orbital phase $\phi$ as described by equation \ref{eq:phi}.

The geometric albedo $p$, which appears in equation \ref{eq:fluxratio}, is the reflectivity of a planet measured at superior conjunction. It is a wavelength dependent, dimensionless number, with values between $0$ and $1$ obtained as ratio between the reflected and incident flux. 

\subsection{Exoplanet's radial velocity} \label{sec:vrad}

The radial velocity equations describing the system are

\begin{equation} \label{eq:rvstars}
    RV_{\rm star} =  \gamma + K_{\rm star} [\cos(\omega + f) + e \cos(\omega)]
\end{equation}
and
\begin{equation} \label{eq:rvstarp}
    RV_{\rm planet} =  \gamma - K_{\rm planet} [\cos(\omega + f) + e \cos(\omega)]
\end{equation}
for the star and the planet respectively. $\gamma$ corresponds to the system's barycenter radial velocity relatively to the Sun, $\omega$ is the argument of periastron, $e$ is the eccentricity, and $f$ is the true anomaly; $K_{\rm star}$ and $K_{\rm planet}$ are the radial velocity semi-amplitude of, respectively, the stellar and planetary orbits and can be computed from the orbital parameters as

\begin{equation}
    K_{\rm star} = \frac{2\pi a}{P_{\rm orb}} \frac{m_{\rm planet}}{m_{\rm star} + m_{\rm planet}}\frac{\sin i}{\sqrt{1-e^{2}}}
\end{equation}
\begin{equation}
    K_{\rm planet} = \frac{2\pi a}{P_{\rm orb}} \frac{m_{\rm star}}{m_{\rm star} + m_{\rm planet}}\frac{\sin i}{\sqrt{1-e^{2}}}
\end{equation}
where $m_{\rm star}$ and $m_{\rm planet}$ are the two masses, of the star and planet, respectively.
The true anomaly $f$, as a function of time $t$ can be computed (see for example \cite{Mortier16}) from

\begin{equation}
    \tan\frac{f}{2} = \sqrt{\frac{1+e}{1-e}}\tan\frac{E}{2}
\end{equation}
where $E$, the eccentric anomaly, can be found solving the Kepler equation

\begin{equation}
    E - e \sin E = 2\pi \frac{t-t_{\rm 0}}{P_{\rm orb}}
\end{equation}
Knowing $\phi$, the planet radial velocity relative to the star can be obtained
as the difference between $RV_{\rm planet}$ and $RV_{\rm star}$ which could be also written as \citep[see][]{Charbonneau99}

\begin{equation} \label{eq:rvstar}
    RV_{\rm planet,star} = - K_{\rm star} \frac{ m_{\rm star} + m_{\rm planet}}{m_{\rm planet}} \sin(2\pi\phi)
\end{equation}
This velocity can be used to compute the Doppler wavelength shift between the stellar and planet spectra at each $\phi$ in equation~\ref{eq:s_ica}.

\subsection{Independent Component Analysis technique} \label{sec:ica}

The Independent Component Analysis (ICA) technique was initially proposed to solve the blind source separation (BSS) problem.
In the BSS problem the task is to separate unknown individual signals (sources or also components) from mixtures of signals without \textit{a priori} knowledge of the mixing process. In ICA, as the name implies, the basic goal is to find a linear transformation in which the underlying components  are statistically as independent from each other as possible. ICA differs from the  more common approach to component separation provided by PCA (Principal Component Analysis) which performs a linear transformation of the data to obtain mutually uncorrelated, orthogonal directions, called indeed  principal components. If the hidden signals under investigation follow Gaussian distributions, uncorrelatedness is equivalent to mutual independence  and algorithms such as PCA are able to separate them. However if the signals follow a non-Gaussian distribution (and most astronomically observed signals are predominantly non-Gaussian), it can be shown that uncorrelated signals are not necessarily mutually independent and hence methods like PCA fail to optimally separate individual components \citep{Waldmann12}. It is actually here where ICA-based separation methods could be of interest.

A comprehensive  description of ICA technique can be found in \citet{Hyvarinen01}. Hereafter we just recall that in ICA the model for the data can be expressed as

\begin{equation}
  \mathbf{X = AS}, \label{eq:icamodel}
\end{equation}
where $\mathbf{X}=(x_1,x_2,\dots,x_m)^T$ are the observed $m$ mixtures, $\mathbf{S}=(s_1,s_2,\dots,s_n)^T$ are the latent variables (i.e. components) that cannot be observed and $\mathbf{A}$ is an unknown constant $m \times n$ matrix, called the mixing matrix. We use bold upper-case letters to denote matrices and lower-case letters (for example $x_1$) to denote, instead, vectors.

The challenge is to estimate the mixing matrix and its
(pseudo) inverse de-mixing matrix, $\mathbf{W = {A}^{-1}}$, without any additional prior knowledge of either $\mathbf{A}$ and 
$\mathbf{S}$.
The estimation of the matrix $\mathbf{S}$ with the knowledge of $\mathbf{X}$ is the linear source separation problem 
and can be achieved by 
maximizing some measure of independence. Several of such measures are used, like kurtosis, negentropy, or mutual information \citep{Hyvarinen01}
and a large variety of algorithms  implementing the above mentioned independence measures (e.g. Second Order Blind Identification,  
SOBI, \cite{Belouchrani97}; Joint Approximation Diagonalization of Eigenmatrices, JADE, \cite{Cardoso93}; fixed-point algorithm, FastICA, \cite{Hyvarinen99} and many others \citep[see][]{Waldmann12}) exists.
\subsubsection{FastICA}
\label{subsub:fastica}

In the present work we decided to use  the FastICA algorithm since it is one of the fastest, most commonly used and it is well documented. It is also available in many programming languages and, in particular, we use the FastICA package for Matlab\texttrademark.
A detailed description of the algorithm can be found in \cite{Hyvarinen99}. Here we just recall its main steps:
\begin{enumerate}
    \item data are centered  (set them at mean zero);
    \item data are whitened (i.e.  $\mathbf{X}$  data are transformed so that the components of the new vector $\mathbf{\tilde{X}}$ are uncorrelated and their variances equal to unity);
    \item the inverse de-mixing matrix $\mathbf{W}$ is estimated;
    \item the mixing matrix $\mathbf{A}$ and the component matrix $\mathbf{S}$ are computed.
\end{enumerate}

The matrix $\mathbf{W}$ is estimated raw by raw with the following iterations scheme:
\begin{enumerate}\renewcommand{\labelenumi}{(\alph{enumi})}
\setlength{\itemindent}{0.5in}
    \item choose initial (random) weight vector $w_{\rm i}$ \label{enum:st1}
    \item let $w^{+}_{i} = E[\mathbf{X}g(w^{T}_{i}\mathbf{X})] - E[g^{'}(w^{T}_{i}\mathbf{X})]w_{i}$, where $g$ and $g^{'}$ are the derivatives of the chosen contrast function (see equation~\ref{eq:gfunc}) \label{enum:st2}
    \item normalize $w^{+}_{i}$ \label{enum:st3}
    \item repeat step (b) and (c) until convergence is achieved
\end{enumerate}
where convergence means that the old and new values of $w_{i}$ point in the same direction, i.e. their dot-product is (almost) equal to 1.

A contrast function $G$ and its first and second derivatives, $g$ and $g^{'}$, are generally used to approximate the negentropy of the system.
In our FastICA package, the choice is restricted to the following functions:
\begin{eqnarray}
    G_{1}(u) & = & \frac{1}{a_{1}}\log \cosh (a_{1}u) \nonumber \\
    G_{2}(u) & = & -\exp(-u^{2}/2) \nonumber \\
    G_{3}(u) & = & \frac{1}{4}u^{4} \label{eq:gfunc}
\end{eqnarray}

The problem of most ICA algorithms is that they are based on methods related to gradient descent where the basic principle is to
start in some initial point, and then make steps in a certain direction until a convergence criterion is met. In case of FastICA sequence
the $w_{i}$ initial vector of weights is generated at random and the algorithm stability is therefore not always deterministic. Moreover, it is  known that equation \ref{eq:icamodel} led to two ambiguities:
\begin{enumerate}
    \item sign  and variances of the independent components could not be determined; \label{enum:amb1}
    \item order of independent components could also not be determined.
\end{enumerate}
Both ambiguities are directly related to the fact that both $\mathbf{A}$ and $\mathbf{S}$ are unknown. Any scalar multiplier in one of the components $s_{i}$ could always be cancelled by dividing the corresponding column $a_{i}$  of $\mathbf{A}$ by the same scalar.  FastICA assumes that each component has unit variance and correspondingly the matrix $\mathbf{A}$ is adapted to take into account this restriction; this solves point \ref{enum:amb1} apart for a sign ambiguity. Concerning the order, positions of independent components can be freely changed without affecting the equation \ref{eq:icamodel}. To show this, it is enough to reshuffle equation \ref{eq:icamodel} by multiplying it with a permutation matrix $\mathbf{P}$ and its inverse as $\mathbf{X = AP^{-1}PS}$. The matrix $\mathbf{AP^{-1}}$ is the new unknown mixing matrix with order of column position changed.
A way to mitigate the FastICA instability and ambiguities  is to run ICA algorithms many times and measure ``somehow'' the stability of the obtained components. This is the purpose of ICASSO, a software package aiming at investigating the relations among estimates from FastICA both programmatically as well as visually.
\subsubsection{ICASSO}
\label{subsub:icasso}

The ICASSO method is described in \cite{Himberg04} and is based on estimating a large number of candidate independent components by running the FastICA algorithm many times, and visualizing their clustering in the signal space. If an independent component is reliable (almost) every run of the algorithm should produce one point in the signal space that is very close to the ‘‘real’’ component. Thus, reliable independent components correspond to clusters that are small and well separated from the rest of the estimates. In contrast, unreliable components correspond to points which do not belong to any cluster.

In ICASSO, independent components estimates can be computed either by randomizing initial condition or by bootstrapping. In the first case, 
FastICA is run $M$ times ($M$, user definable number of iterations) on the same data $\mathbf{X}$, but starting each time with a new random initial condition; in the second case, the $M$ runs are performed keeping the same initial condition, but the data are re-sampled by bootstrapping them every time. 
Obtained estimates at each run are afterwards clustered according to their mutual similarities (agglomerative clustering with
average-linkage criterion is actually used), visualized in a 2-D plot and made available through dedicated application programming interface (API) for successive analysis.

In ICASSO the clustering performed by the software package is already capable to group similar components obtained over the many, user-defined, runs, based on the absolute value of their mutual correlation coefficients (for details see \cite{Himberg04}). 
Despite the ambiguities described in \ref{subsub:fastica}, the similarity criterion allows to integrate estimates of all the runs in a new single
estimate, called in ICASSO, ''centrotype''. Basically it is the point in the cluster that has the maximum sum of similarities (as measured by correlation coefficients) respect to the other points in the cluster. The obtained estimates therefore not only can be considered ''reliable'', but also ''improved'', if compared with a single FastICA run.

 The coupling of the ICASSO package  with the FastICA one in the Matlab\texttrademark environment was also one of the driver to select specifically FastICA algorithm among the various ICA algorithms described in the literature.
\section{Method description and validation} \label{sec:simul}

In the present work we want to apply ERLICA to decompose an observed spectrum produced by an exoplanetary system (star plus one planet) in individual components. The observed mixture $\mathbf{X}$ matrix is, in this case, composed by  the $x_j$ (observed) spectra, where $x_j = x_j(\lambda_0, \cdots ,\lambda_l)$ represent one 
realization out of $m$ (observed) spectra, whereas the 2 individual components that we want to extract, $s_{1}$ and $s_{2}$, are the stellar and exoplanet spectra. Note that for non-overcomplete sets the number of (observed) spectra should be equal to the individual source signals (i.e. $m=2$ for our specific case). For overcomplete sets (i.e. more data than source signals) dimensionality can be reduced given some selection criteria (e.g. signal-to-noise ratio, in case of stellar spectra, or others) thus reducing the set to the non-overcomplete case.

Before applying our ERLICA method to real observations we decided to test it on artificial (simulated) data  in order to be able to assess its effectiveness and validity. Following the notation of equation \ref{eq:fluxratio} we will identify hereafter $s_{1}$ as  $F_{\rm star}$,  $s_{2}$ as  $F_{\rm planet}$ and individual (observed) spectra as $F_j(\lambda)$.
The validity of using ICA relies on the assumption that $F_{\rm star}$ and  $F_{\rm planet}$ are independent while  its effectiveness should be related to the $SNR$ (Signal-to-Noise Ratio) of the input data.  
Statistical independence of $F_{\rm star}$ and  $F_{\rm planet}$ means that, at each wavelength, $F_{\rm planet}(\lambda_i)$ does not depend on 
$F_{\rm star}(\lambda_i)$. In our case,  this can be  expected due to the relative Doppler shift of the two spectra and to the modulation  introduced by the albedo.
A  qualitative indication of independence can be inferred by looking at the soundness of the results of applying ERLICA on simulated data. In fact,  only if the independence assumption is valid, ERLICA would be able to provide reliable results.

Since, in  the case of real statistical independence, the joint pdf (probability density function) is factorizable into the product of the single marginal pdfs,  a further quantitative indication  can be assessed   by checking  the  similarity between  the frequency distributions  $f(F_{\rm star}|F_{\rm planet})$ and  $f(F_{\rm star}) \times f(F_{\rm planet})$. Actually, we found  that their average difference, $<f(F_{\rm star})|F_{\rm planet})-f(F_{\rm star} \times f(F_{\rm planet})>$, is equal to zero within $2\sigma$ (see results in Section\,\ref{sec:simul1}) thus confirming the ``practical'' independence of $F_{\rm star}$ and  $F_{\rm planet}$.

It is important to note that in general the observed signal is a sum of the two astrophysical signals and of various, even complex, sources of noise and it is characterized by its $SNR$. 
To take into account in our simulation the $SNR$ role, we mimic, without losing in generality,  only the case of pure Gaussian noise ignoring  systematics introduced by several causes like instrumental signature, stellar activity, telluric fluctuations etc. \citep[see discussion in][]{Waldmann13}.
In fact at the level of simulation, adding additional non-Gaussian systematic components will not add new information to the problem of components separations since, due to its nature, ICA will be anyway able to disentangle them. On the contrary, adding specific systematic will require to tie simulations to a specific instrument (which we would like to avoid to stay general) and force us to use a more complex observed matrix (with as many rows as the component we would like to search for).

\subsection{Simulation of an exoplanetary system: the case of 51\,Peg\,b} \label{sec:simul1}
To represent a possible real case, we decided to simulate the 51\,Peg~+~51\,Peg\,b planetary system,  Indeed, this is also one of the first systems where reflected signal from an exoplanet has been detected at a significance of $3\sigma_{noise}$ \citep{Martins15}. To properly characterize the system we use the parameters summarized in Table~\ref{tab:Peg51b}. 

The model of the composite spectrum at different orbital phases is obtained, similarly as it was done in \cite{Borra18}, by adding to a stellar spectrum the same spectrum Doppler shifted and modulated by:
\begin{itemize}
    \item a phase-dependent function
    \item the predicted geometric flux ratio of the planet relative to the star
    \item a reflecting albedo
\end{itemize}
as described in sections  \ref{sec:phase} and \ref{sec:vrad}.

\begin{deluxetable*}{ccl}[h!]
\tablecaption{Nominal orbital parameters for 51\,Peg + 51\,Peg\,b system \label{tab:Peg51b}}
\tablecolumns{3}
\tablewidth{0pt}
\tablehead{
\colhead{Orbital parameter} &
\colhead{Value} &
\colhead{Reference}
}
\startdata
$t_{0}$ & 2456021.256 JD & \citet{Martins15}\\
$P_{orb}$ & 4.231 day & \citet{Martins15} \\
$m_{star}$ & 1.12 $M_{sun}$ & \citet{Fuhrmann97} \\
$m_{planet}$ & 0.46 $M_{J}$  & \citet{Martins15}\\
$a$ & 0.052 AU& \citet{Martins15} \\
$i$ & 80 deg & \citet{Martins15}\\
$\omega$ & 0 & \citet{Martins15}\\
$\gamma$ & -33.152 & \citet{Martins15}\\
$R_{star}$ & 1.20 $R_{sun}$ & \citet{Fuhrmann97} \\
$R_{p}$ & 1.9 $R_{J}$ & \citet{Martins15}\\
\enddata
\end{deluxetable*}

\begin{figure}[ht!]
\plotone{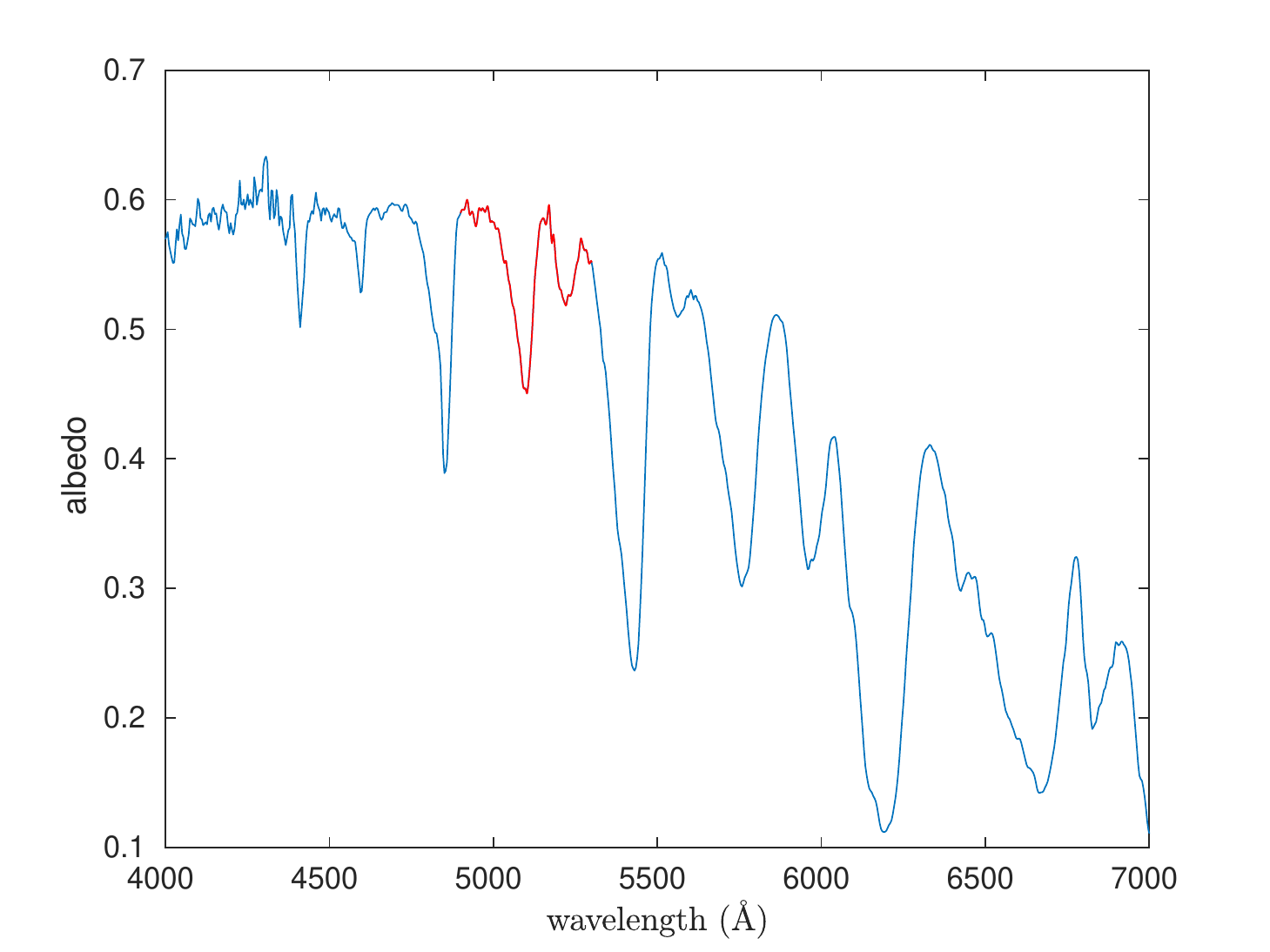}
\caption{Full-disk albedo  of Neptune as obtained by \cite{Karkoschka98};
red portion of the albedo highlights the wavelength range used in the paper\label{fig:albedo}}
\end{figure}

In order to simulate the composite spectrum, we used as $F_{\rm star}(\lambda)$  the synthetic spectrum  of 51\,Peg  computed by means of 
SPECTRUM v.276e \citep{Gray94} starting from an ATLAS12 atmosphere model \citep{Kurucz05} 
at $T_{\rm eff} = 5787$\,K, log\,$g = 4.45$\,dex, metallicity $0.15$\,dex and microturbulence $0.85$\,km\,s$^{-1}$ \citep[data taken from][]{Valenti05}.
The synthetic spectrum was broadened to take into account the rotational velocity of $2.6$\,km\,s$^{-1}$, macroturbulence velocity of $3.95$\,km\,s$^{-1}$ and degraded 
at the resolution of ESO HARPS spectrograph (R $\approx$ 115,000).
The same synthetic spectrum, but Doppler shifted, was used to compute the $F_{\rm planet}(\lambda)$. The planetary albedo $p(\lambda)$ was assumed to be equal to the Neptune one given by \cite{Karkoschka98} and it is shown in figure~\ref{fig:albedo}.  More recent measurements by \citet{Madden18} are also available in the literature, but  we decided to use the ones by \cite{Karkoschka98} due to their more extended wavelength coverage towards the blue.
We decided to use the Neptune albedo because in the wavelength range of our simulation it shows prominent features with variations on the order of $\pm 30\%$.
The goodness of our method in extracting planet features can be in this way better assessed. Actually at the level of simulation what is important is to show the ability to recover the planet signal given in input; in principle, without any loss in generality, we could even not use any real, observed albedo, but simply adopt for $p(\lambda)$ a monotonic function.     

In our simulation the spectra are limited to the $4900 - 5300$ \AA\,wavelength range and have a fixed step of $0.005$\AA. The wavelength range has been chosen in a spectral region where there is a possibility of achieving good signal-to-noise ratio in the case of  real ground-base observations avoiding  regions of telluric contamination and the  presence of 
strong lines (e.g. hydrogen ones), yet, keeping the computational time reasonable.  

To create the synthetic observed mixture $\mathbf{X}$ (composed by different $F_{j}(\lambda)$), to be analyzed subsequently by ICA, one should note that it is not possible to use $F_j(\lambda)$ for all different $\phi$ simultaneously. In fact, the contribution of the $F_{\rm planet}(\lambda,\phi)$ in each $F_j(\lambda)$, at different phase angles, would be seen by ICA as a different independent component (due to the different radial velocity shift). This will lead to an attempt to disentangle an observed mixture $\mathbf{X}$ composed, say, by $m F_{j}(\lambda)$ in $n=m+1$ components i.e. $m$ $F_{\rm planet}(\lambda,\phi)$ and one $F_{\rm star}$, which has of course, no solution. We actually need to inject in ICA an observed mixture such that the dimensionality is correctly preserved. For our specific exoplanetary case this can be achieved, in the most simple case, by using  pair of spectra $F_{j}(\lambda)$ where in one of them the exoplanet reflected spectrum is not present. The  observed mixture $\mathbf{X}$ assumes in this case the form:

    \begin{align} \label{eq:x_ica}
     \mathbf{X}  =  \begin{pmatrix} F_{1}(\lambda) \\ F_{2}(\lambda) \end{pmatrix} 
    \end{align}  

where:

  \begin{eqnarray} \label{eq:s_ica}
      & F_{1}(\lambda)~~~ & =  F_{\rm star}(\lambda) \nonumber \\
     & F_{2}(\lambda,\phi) & =  F_{\rm star}(\lambda)+F_{\rm planet}(\lambda,\phi) = F_{\rm star}(\lambda) + p(\lambda)\left[\frac{R_{\rm planet}}{a}\right]^2g(\alpha)  F_{\rm star}\left(\lambda\left[ 1+\frac{RV_{\rm planet,star} }{c}\right]\right) 
  \end{eqnarray} 

In equation~\ref{eq:s_ica}, $\phi \in [0,1]$ is the usual orbital phase (see equation \ref{eq:fi}) and $RV_{\rm planet,star}$ is the radial velocity of the planet relative to that of the star (as derived in equation \ref{eq:rvstar}). $F_{1}(\lambda)$ thus effectively represents a spectrum 
obtained at the planet occultation ($\phi=0.5$)  if it is an eclipsing system, or at the inferior conjunction ($\phi=0.0$) if not, i.e. when no star light is reflected by the planet towards the observer.  $F_{2}(\lambda)$, instead, contains also the contribution reflected by the planet at the corresponding phase $\phi$.
As discussed in  Section\,\ref{sec:simul} to check the independence of $F_{\rm star}$ and  $F_{\rm planet}$ we computed $f(F_{\rm star}|F_{\rm planet})$ and  $f(F_{\rm star}) \times f(F_{\rm planet})$ 
and we found an average difference, $<f(F_{\rm star})|F_{\rm planet})-f(F_{\rm star} \times f(F_{\rm planet})>$, equal to $2.7 \times 10^{-6} \pm 2.1 \times 10^{-6}$. Such a low difference suggests that the two  spectra are independent and thus we can be confident that ERLICA will be able to extract the two searched  components, i.e. the stellar spectrum and the reflected  light from exoplanet properly modulated by the introduced albedo.

In order to simulate a realistic case, as discussed in section \ref{sec:simul}, we added to both spectra realizations, by using Matlab\texttrademark function \textit{normrnd}, a gaussian noise 
with mean zero and  standard  deviation $F(\lambda)/SNR$ (where $SNR$ is a parameter to specify the signal-to-noise ratio of the $m F_{j}(\lambda)$ spectra in the wavelength range of interest). 
Expressing the noise via $SNR$ is convenient and  allows us to estimate the validity of the obtained results as a function of varying $SNR$.

\subsubsection{Extraction of ICA components}\label{subseq:extraction}

\begin{figure}[ht!]
\plotone{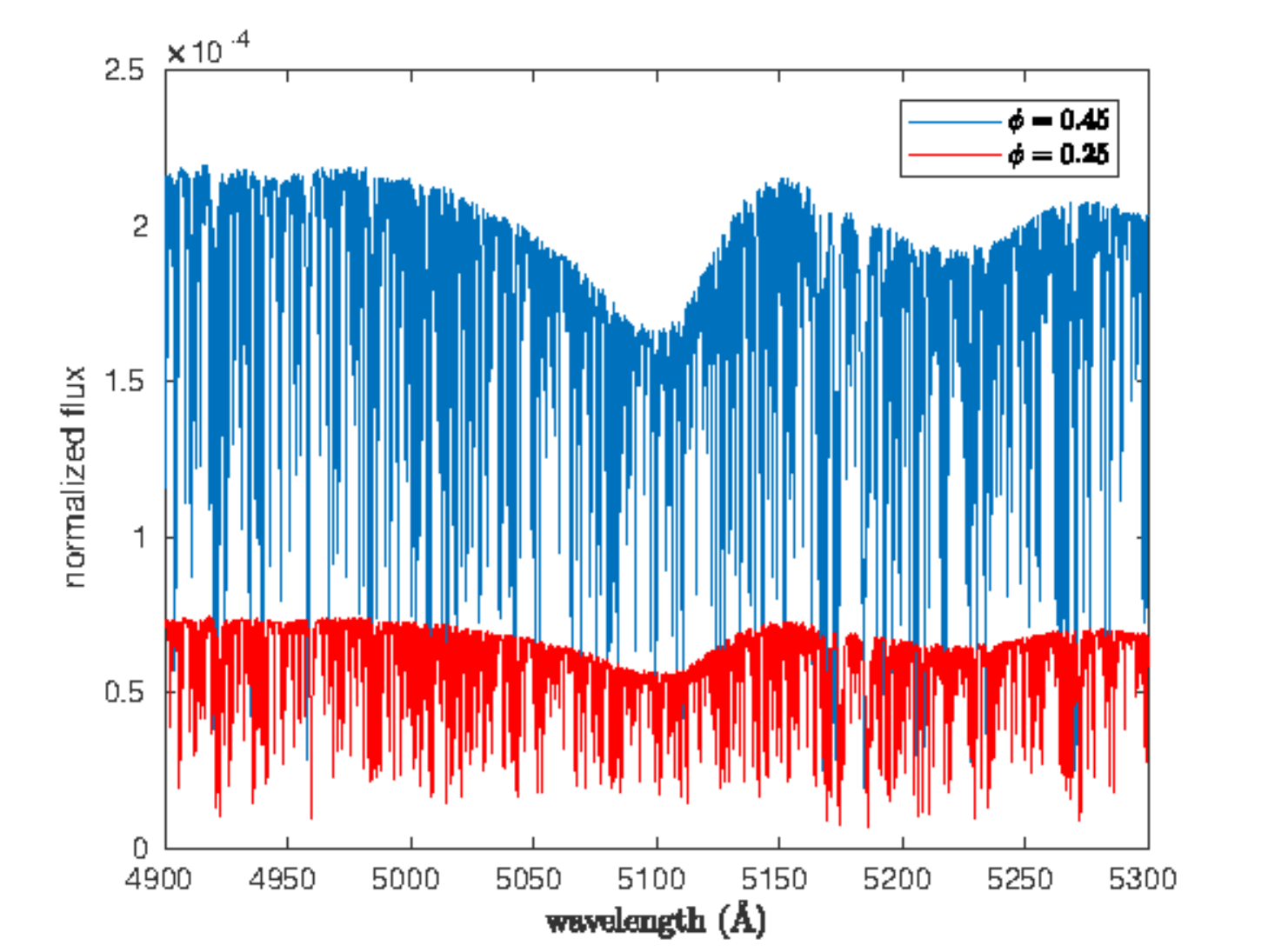}
\caption{Simulated reflected spectrum of 51 Peg\,b at two different orbital phases $\phi$: $\phi = 0.45$ (blue line) corresponding to the case of maximum reflection (planetary disk almost fully illuminated) and at $\phi = 0.25$ (red line) corresponding to the case of maximum separation in radial velocity.\label{fig:reflected}}
\end{figure}

In order to assess the behaviour of ICA in disentangling the system individual components we started by applying ERLICA, i.e. FastICA+ICASSO,  on the composite spectra obtained when 51 Peg and 51 Peg b are at the two phases presented in figure~\ref{fig:reflected}. These configurations represent somehow the two competing ''extreme'' cases: the case of maximum available flux signal and still different planet and stellar radial velocities   vs the maximum shift in radial velocity  but a lower phase dependent flux ratio.  The blue spectrum in the figure corresponds to the case of the exoplanet disk almost fully illuminated, $\phi=0.45$, but with a minimal separation in radial velocity ($RV_{\rm planet,star} \approx 40.7~$km\,s$^{-1}$). In this case, the phase dependent flux ratio (see equation~\ref{eq:fluxratio}) amounts  to $2.9 \times 10^{-4}$. The red spectrum, on the contrary, represents the case of maximum shift in radial velocity ($\phi=0.25$, $RV_{\rm planet,star} \approx 131.7~$km\,s$^{-1}$), but with a disk only partially illuminated leading to a phase dependent flux ratio of $9.7 \times 10^{-5}$.

As a starting point a $SNR=50,000$ has been adopted to limit the influence of the noise and highlight the capabilities of the method. 

In extracting the two independent components we have to face the two ambiguities described in Section~\ref{sec:ica}.
The first one, i.e. the sign of each component, can be solved taking into account that each $F_{2}(\lambda,\phi)$ must be the sum of two positive quantities. Therefore we determined the sign of each disentangled components by taking into account the sign of the corresponding element of the matrix $\mathbf{A}$.

The second ambiguity, i.e  the order of the components, was resolved by
comparing each of them with  one of the two input spectra (each of them being dominated by the stellar signal) via cross-correlation. The output component with the highest value of the cross correlation is then labeled as component~1 and identified as the ICA estimate of $F_{\rm star}$.

Setting-up the simulation is however not enough:  we need a way to judge and assess, with the results in hand, 
the ''goodness'' of ICA in extracting the two signals. 
We decided, therefore, to search for a quantitative, mathematical, estimator. A 
first attempt led us to calculate simply the standard deviation of the differences between the extracted components and the
input $F_{\rm star}$ and  $F_{\rm planet}$.  However,  this estimator cannot be use in the case of real 
observations, since there will be no ``a-priori known''  $F_{\rm star}$ and  $F_{\rm planet}$ signals to be compared with. 
Therefore, we decided to use an  approach similar to that one shown in \cite{Borra18}, i.e. to use, as a more reliable and feasible estimate, the peak intensity of the auto-correlation function of each obtained component. In fact, the higher this peak is, the higher  the strength of the detected signal is  as discussed in Section\,4.3 of \citet{Borra18} who adopted a similar approach to assess the detection of a planetary signal with their ACF method.

Therefore, we autocorrelate  each of the two found components extracted by applying ERLICA on the composite spectra. The ACFs are computed, after resampling the components on a velocity scale, by relatively shifting them at steps, lag,  using a constant velocity increment. The ACF profile is, by construction, always symmetrical and well centred at $0$ position.  Actually, correlating a signal by itself always produces a peak centered at $0$ and equal to $1$ after normalization. If there is a signal, i.e. a  detection, it can be inferred from the ACF increase with respect to that one obtained if only noise is present at $X$ values (shifts) equidistant and close to $0$ both on negative and positive side.  Therefore, to use the ACF as an estimator we removed  the autocorrelation peak at 0, which is, for our purposes, not meaningful and substitute it with a value given by  interpolation (usually with a gaussian fitting) of the neighborhood values.
The peak intensity of such modified profile can be used as a robust estimator of the detection level when compared with the same quantity obtained in the case of an ACF computed  when   a detection  is not possible.
No-detection can be simulated by injecting in ERLICA  two  spectra $F_1(\lambda)$ and $F_2(\lambda)$ computed both at $\phi\simeq0.0$ thus using $F_{\rm Planet}=0$ in equation \ref{eq:s_ica}.
In the following, we  use the ratio of the peak of the ACF of each $i$ component over the no-detection ACF peak value to define the detection significance, $D_i$, and declare that there is a detection, if  $D_i$ is larger than $3$.

In summary the main steps of the simulation are the following:
\begin{enumerate}\renewcommand{\labelenumi}{(\roman{enumi})}
\setlength{\itemindent}{0.25in}
\item create $F_{1}(\lambda)$ and $F_{2}(\lambda)$ according to equation\,\ref{eq:s_ica} for a chosen $\phi$ value;
\item add to $F_{1}(\lambda)$ and $F_{2}(\lambda)$ a Gaussian noise to mimic spectra with different $SNR$ values;
\item create the observed mixture $\mathbf{X}$ and apply ERLICA;
\item fix signs and order of the extracted components;
\item compute ACF of each component after re-sampling  at steps of constant velocities;
\item compute detection significance $D_1$ and $D_2$ values.
\end{enumerate}

\subsubsection{Contrast function selection and $SNR$ effect}  \label{sec:gfunc}

 FastICA algorithm implementation allows the user to choose among three different contrast functions (see equation~\ref{eq:gfunc}). The theoretical analysis that led to the adoption of the aforementioned contrasts function can be found in \cite{Hyvarinen99}, here we want just to recall the main conclusions given there:
\begin{enumerate}
    \item $G_1(u)$ is a good general-purpose contrast function;
    \item $G_2(u)$ may be better when robustness is very important;
    \item $G_3(u)$ is not recommended in case of presence of outliers.
\end{enumerate}

To chose the most appropriate contrast function for our scientific case we applied the ERLICA approach as described in section~\ref{sec:simul1} by using all the three contrast functions and varying $SNR$  on the simulated 51 Peg + 51 Peg b planetary system with $\phi=0.45$. $SNR$ has been varied in the range $[1000-50\,000]$ to span a noise amplitude interval from ten times to one fifth of the planet flux. For each value of $SNR$ we computed the detection significance $D_i$ of the disentangled components. 

Figure~\ref{fig:gfunc} shows the results.
Some points can be highlighted: 
\begin{figure}[h!]
\plotone{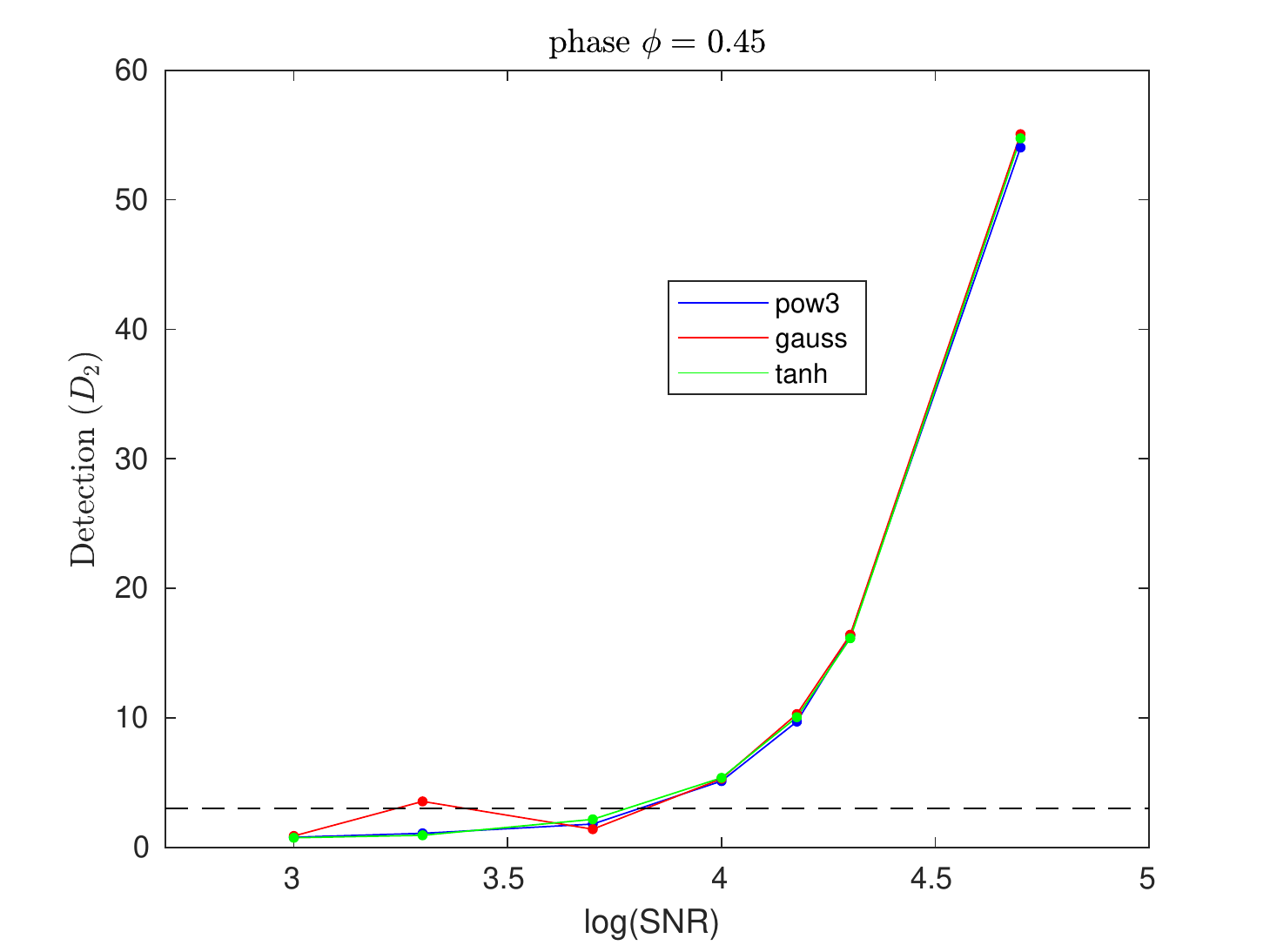}
\caption{Trend of the detection significance for the second component as a function of the logarithm of SNR; different colors represent different adopted contrast functions while the dashed line corresponds to the detection threshold $D_2=3$ (see text)\label{fig:gfunc}}
\end{figure}
\begin{figure}[htbp]
\plotone{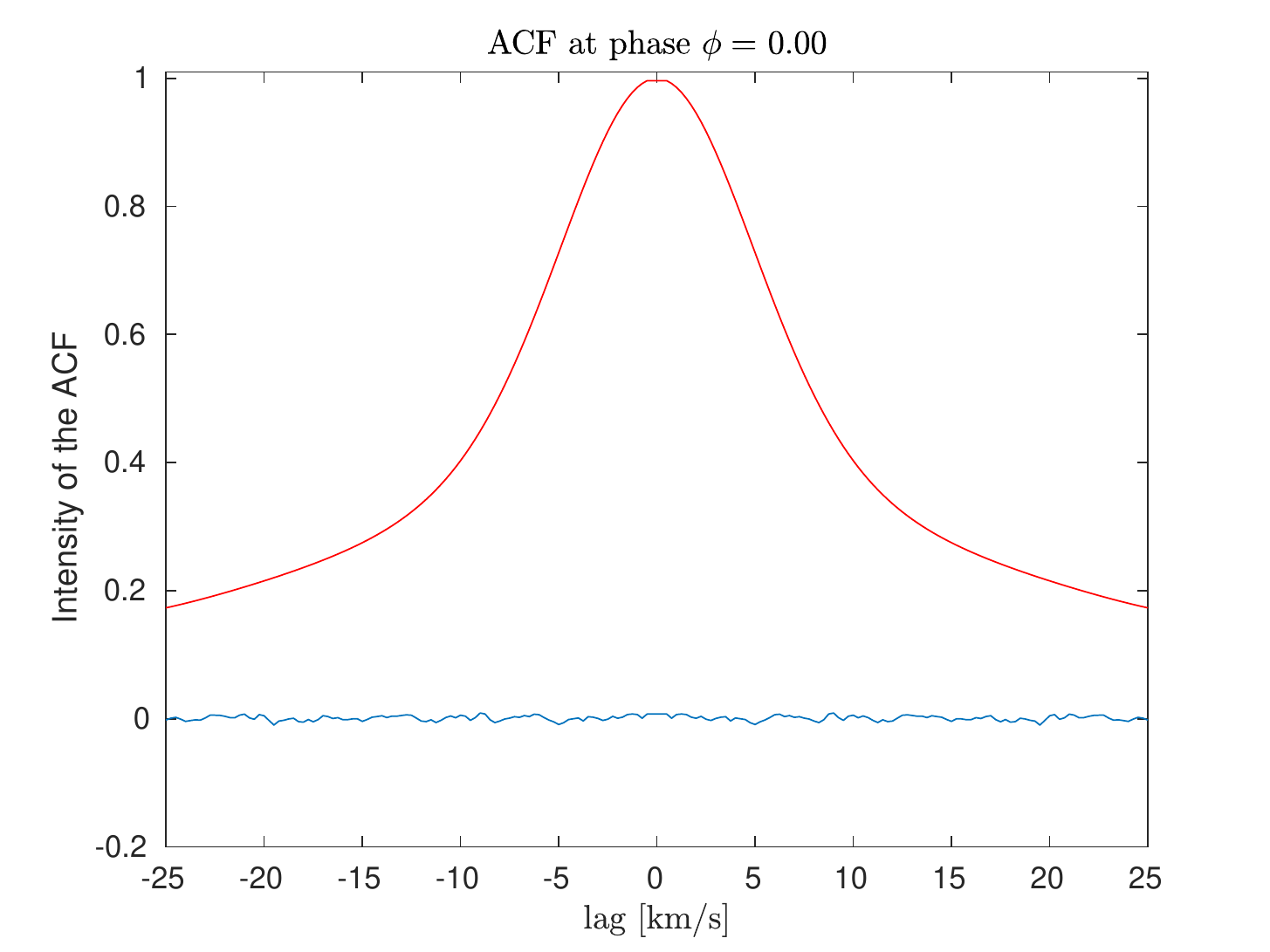}
\caption{ACF profiles for the detected first component (red) and no-detected second one (blue), see text\label{fig:acf_000}}
\end{figure}
\begin{figure*}[htbp]
\gridline{\fig{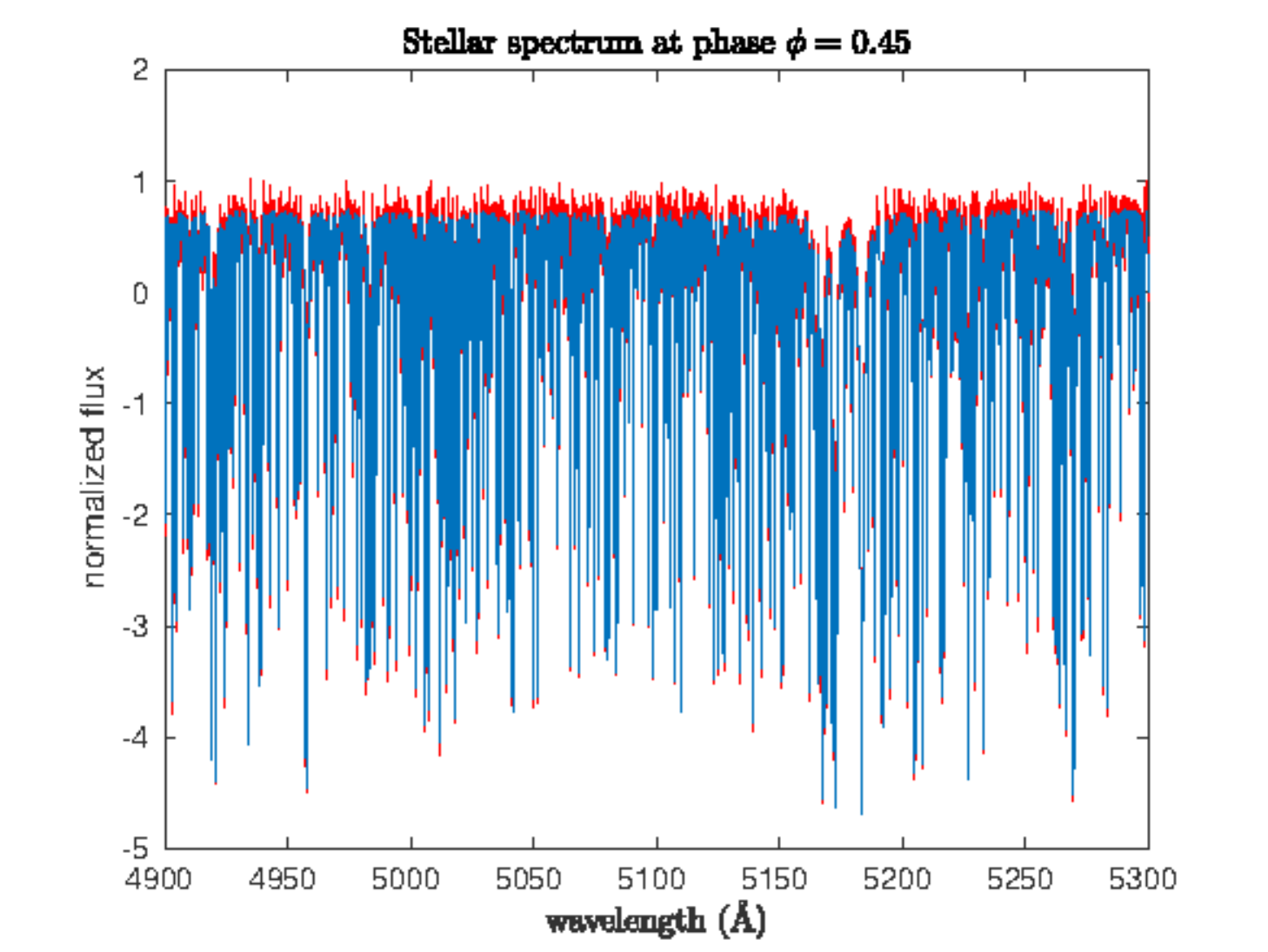}{0.4\textwidth}{(a)}
          \fig{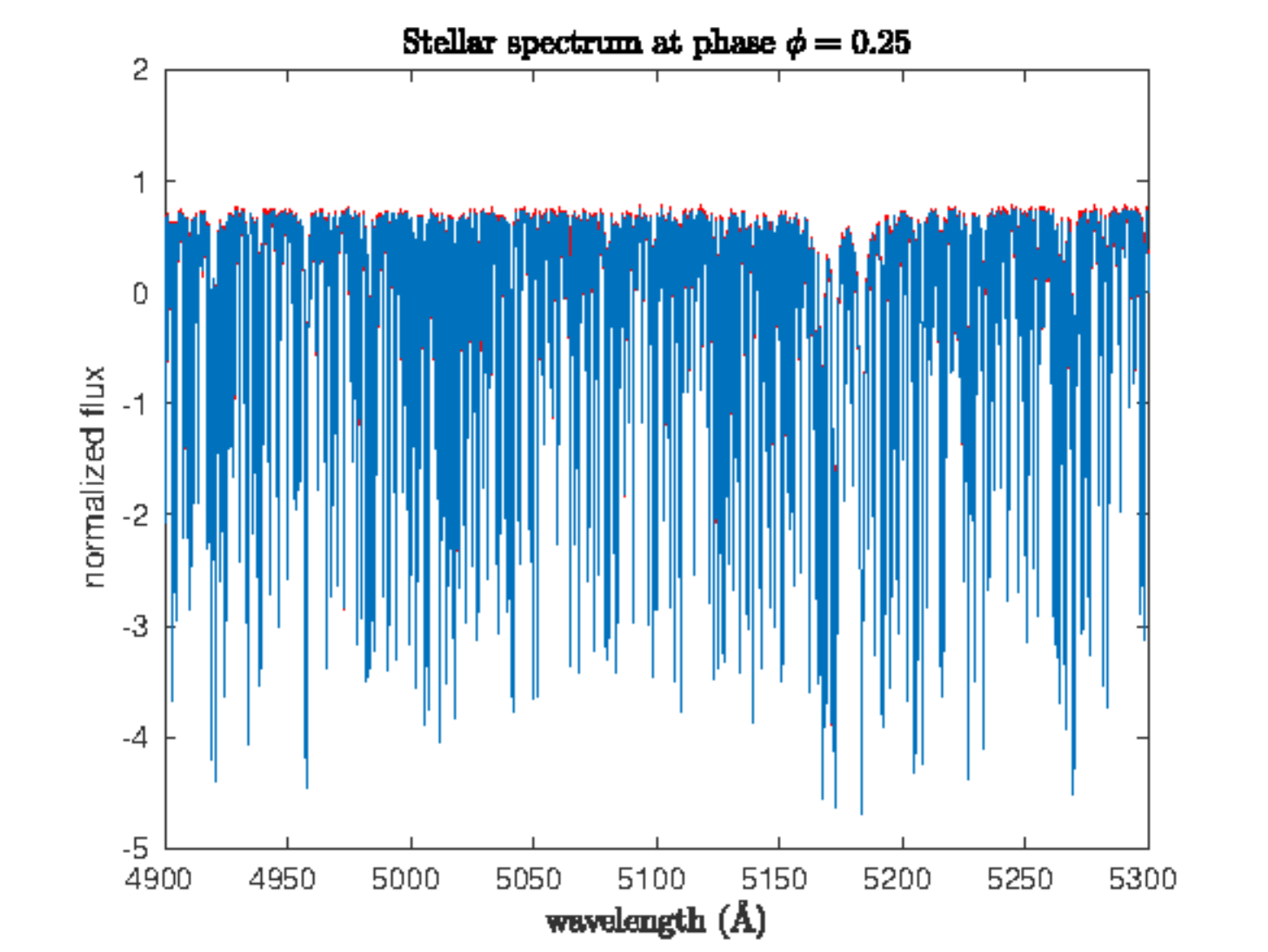}{0.4\textwidth}{(b)}}
\gridline{\fig{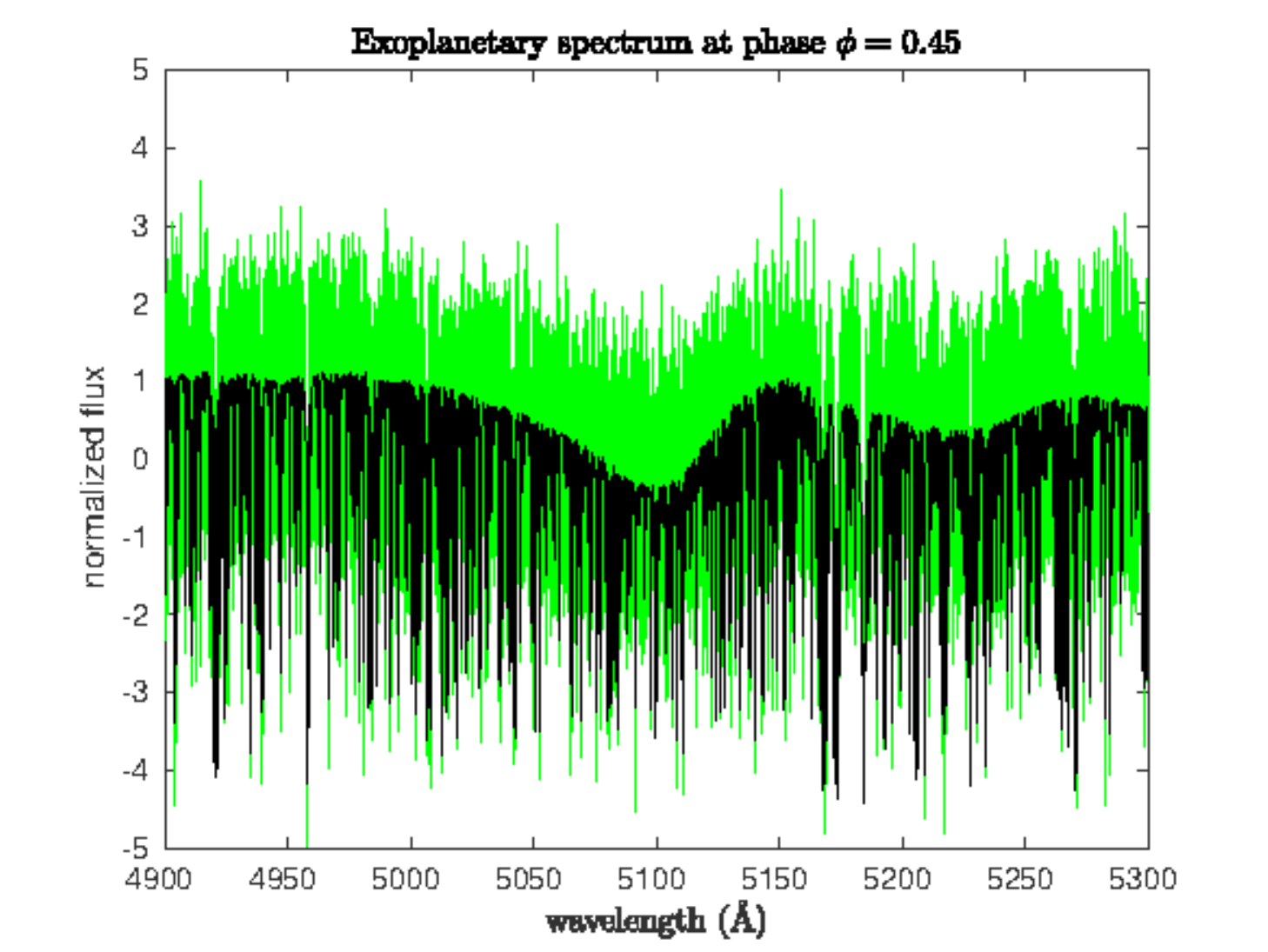}{0.4\textwidth}{(c)}
          \fig{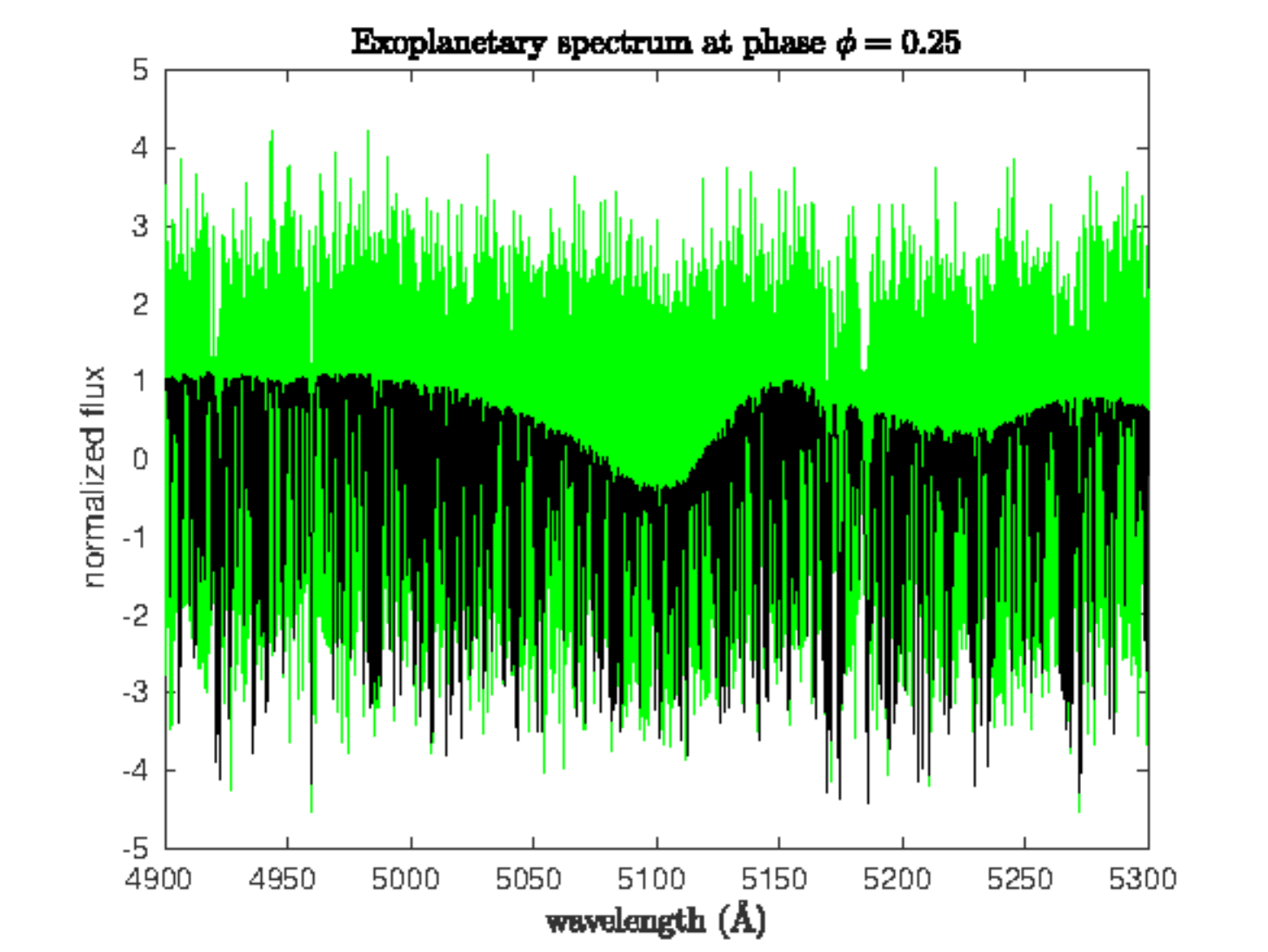}{0.4\textwidth}{(d)}}
\gridline{ \fig{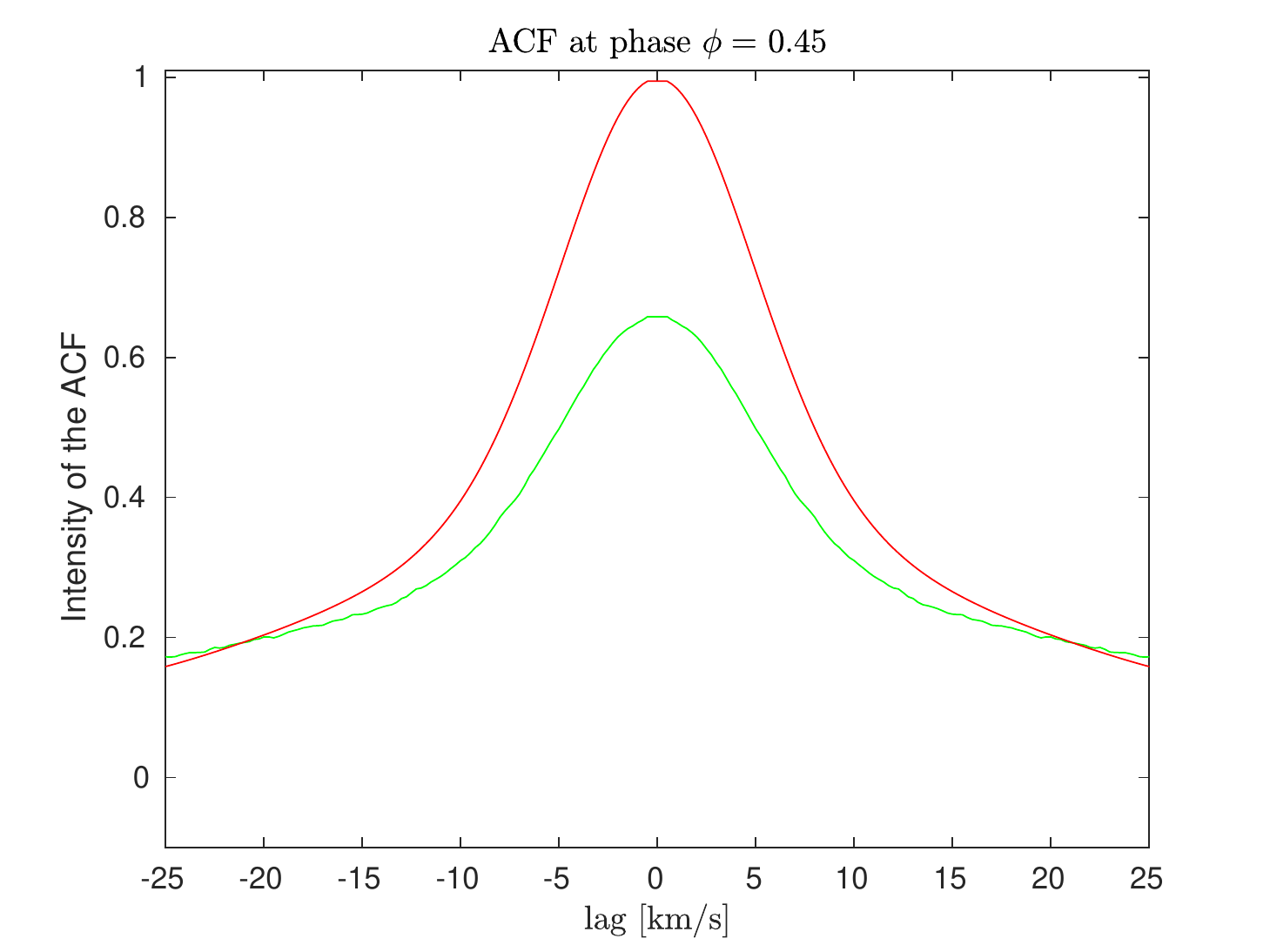}{0.4\textwidth}{(e)}
          \fig{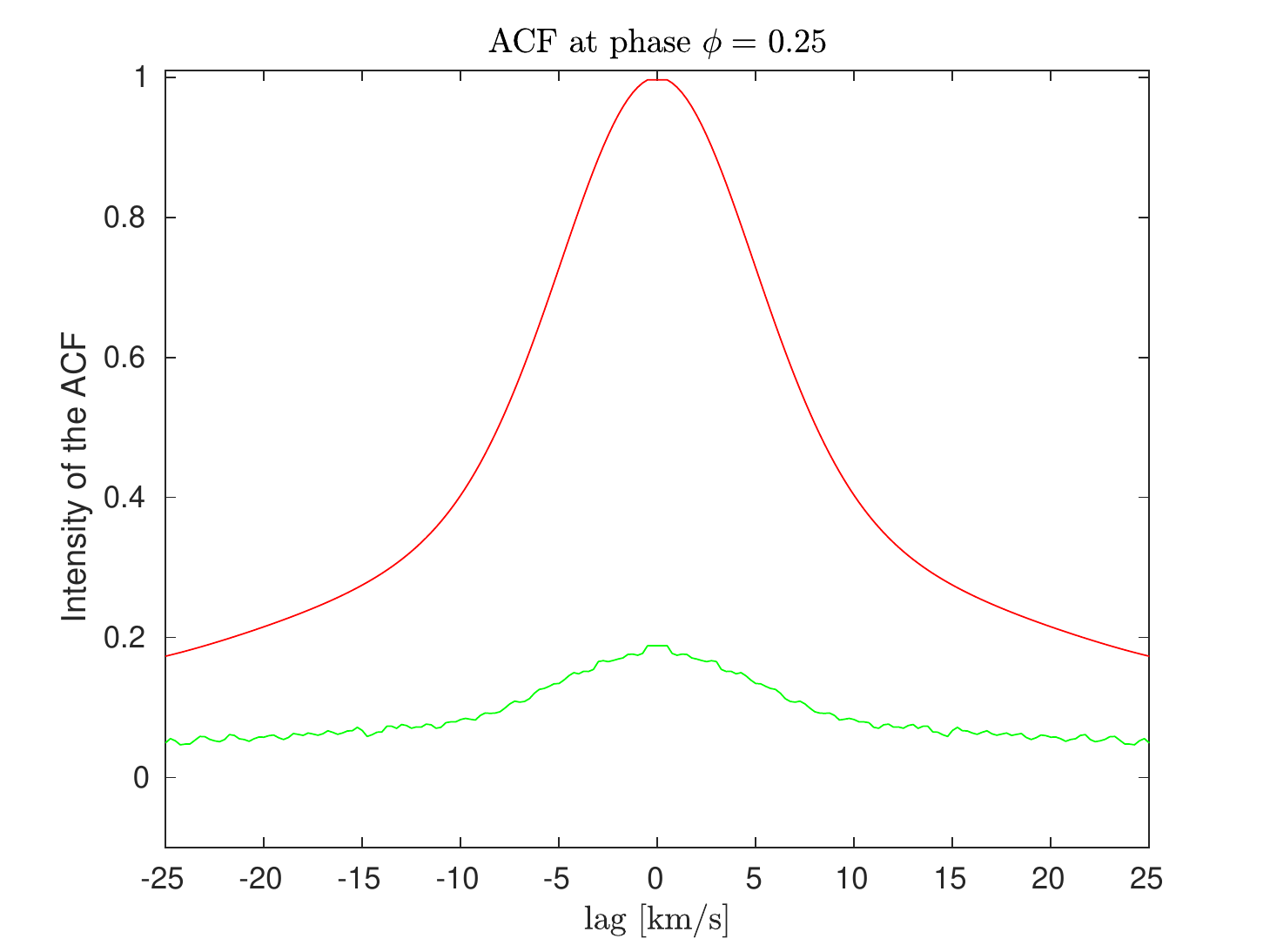}{0.4\textwidth}{(f)}}
\caption{ICA results for $\phi = 0.45$ and $\phi = 0.25$, and $SNR=50,000$. Panels (a) and (b): Detected first component (red) with superimposed the input star spectrum (blue); panels (c) and (d): Detected second component (green) with superimposed the input
exoplanet reflection spectrum modulated by albedo (black); panels (e) and (f): ACF profiles of the first (red) and  second (green) detected components \label{fig:ica_045}}
\end{figure*}
\begin{itemize}
    \item detection level increases with increasing SNR, as expected, independently of the adopted contrast function;
    \item  all the three contrast functions led to almost equivalent results, but for $SNR=5000$.
    The $G_2(u)$ ''gauss'' contrast function shows a non-monotonic behaviour for  $SNR<10000$ which
    seems to indicate a greater sensitivity to the FastICA initial conditions;
    \item in general a $SNR>5000$ is required for a reliable detection ($D>3$). This $SNR$ limit corresponds to the case of a noise amplitude comparable to the planet signal one.
\end{itemize}
\begin{figure*}[htbp]
\gridline{\fig{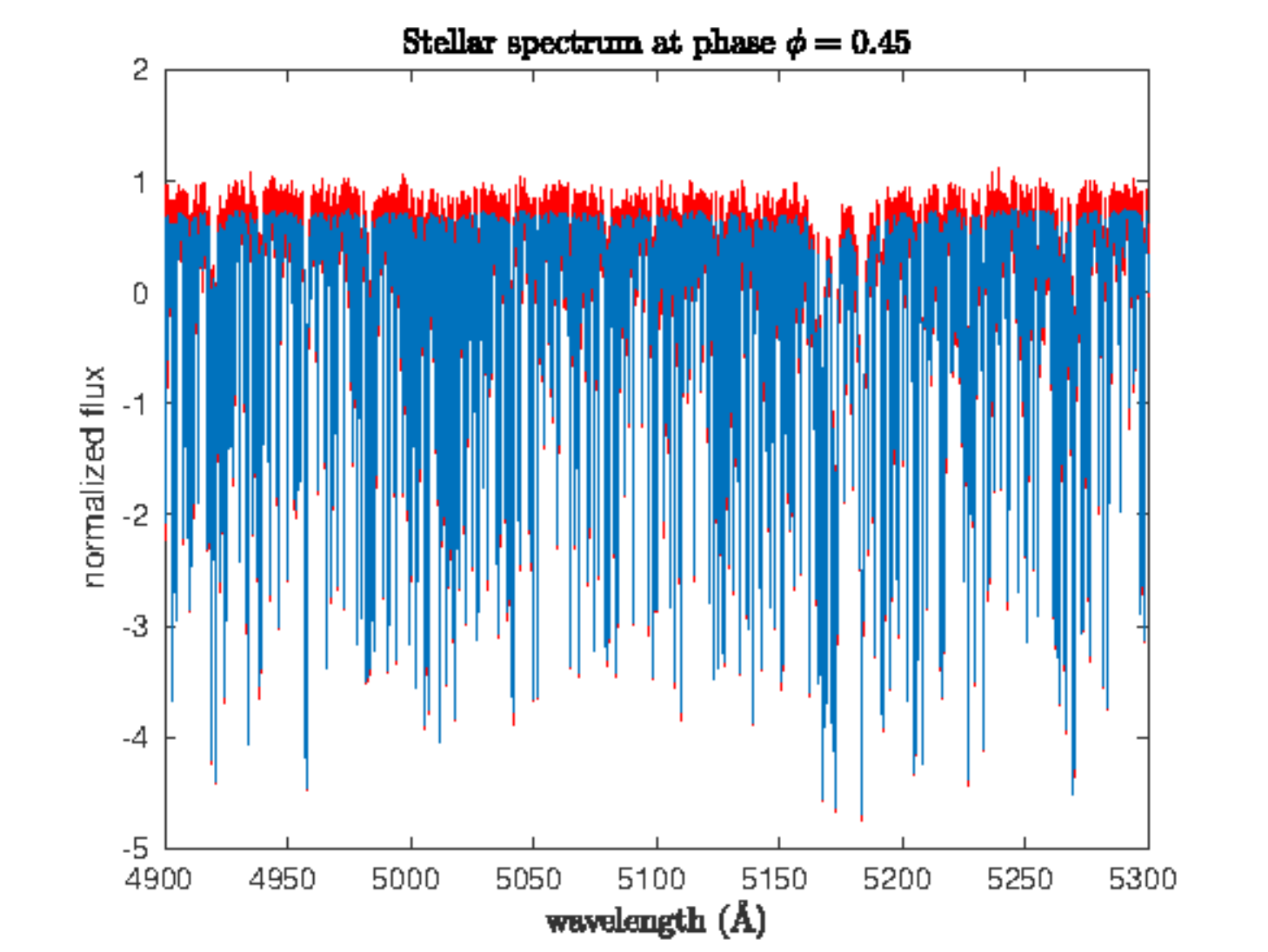}{0.45\textwidth}{(a)}
          \fig{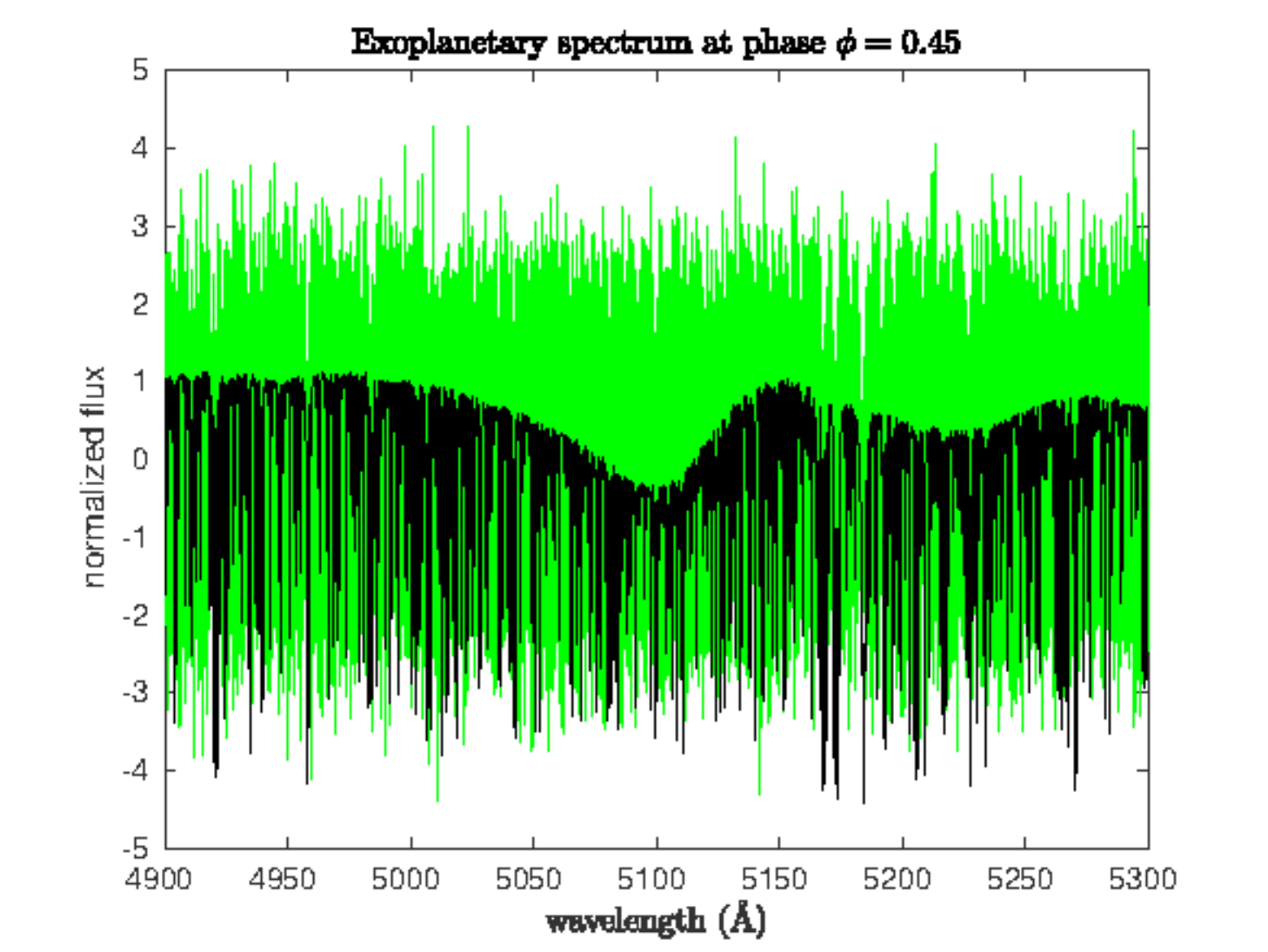}{0.45\textwidth}{(b)}}
\gridline{       
          \fig{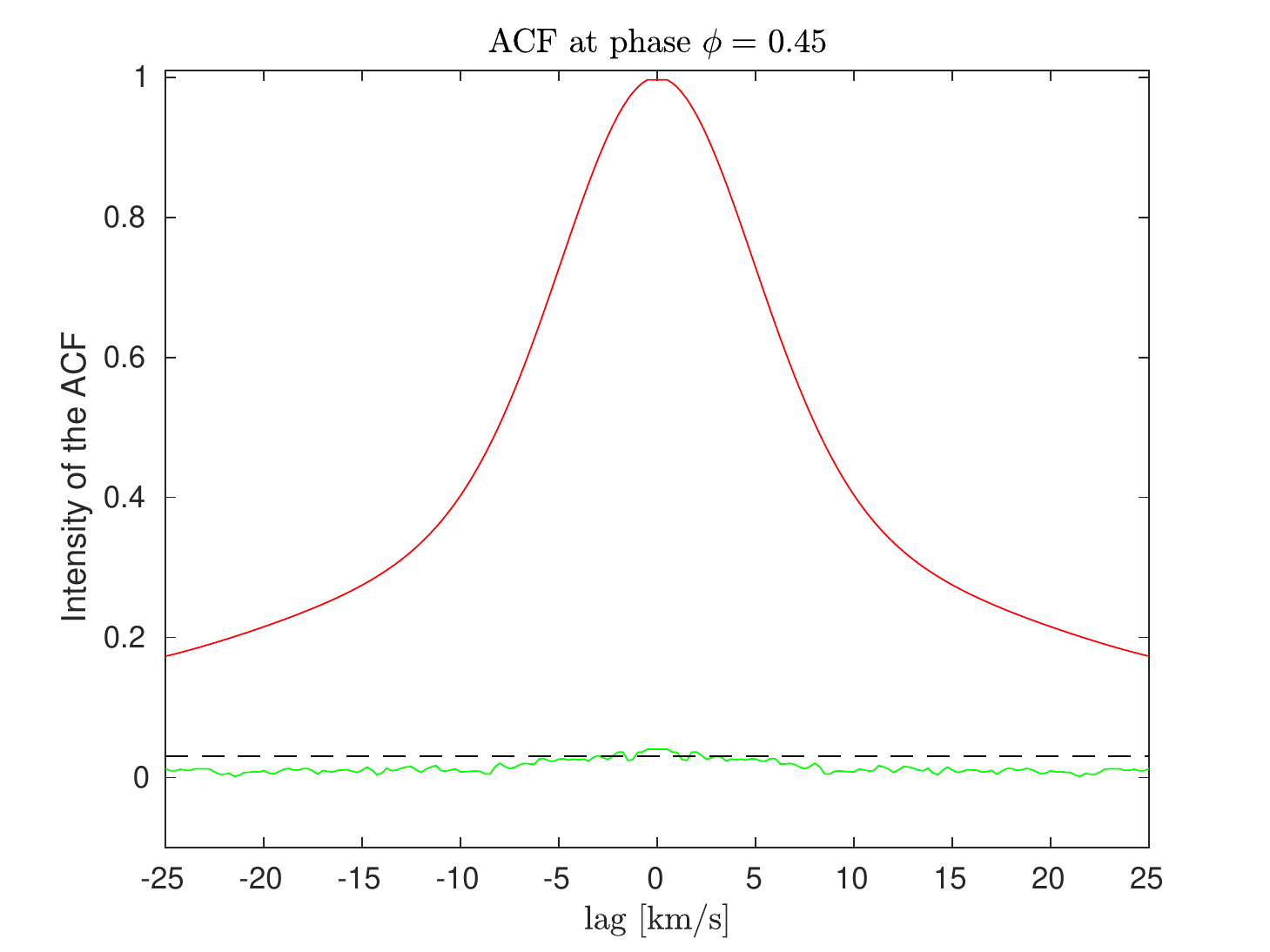}{0.45\textwidth}{(c)}
          }
\caption{As Figure~\ref{fig:ica_045} for $\phi = 0.45$ and $SNR=6,500$. The dashed line corresponds to the detection threshold $D_2=3$ \label{fig:ica_045_7000}}
\end{figure*}

Taking  into account the main conclusions in \cite{Hyvarinen99} and considering that $G_2(u)$ performs, sometimes, slightly better than the other contrast functions going towards low $SNR$ we decided to use the $G_2(u)$ ''gauss'' contrast function  in our ICA analysis. It is worthwhile  noticing that, as shown in our simulations 
and highlighted in \cite{Hyvarinen99}, this choice of the contrast function is not really 
critical in the sense that any of the contrast estimators in FastICA framework works well for (practically) any distributions of the independent components (contrary to what happens in other ICA algorithms).

\subsubsection{Simulation results}\label{subsec:sim_res}

First of all, we apply ERLICA at the case $\phi=0$ where no detection of the second component is expected. The corresponding ACF of the two components profiles are shown in figure~\ref{fig:acf_000}. 
In this case, ERLICA is clearly able to retrieve the first component, which has a well peaked ACF, whereas the ACF of the ''second'' one (actually ERLICA retrieves only noise) is flat and with a maximum value of about $0.01$. 

Then, we repeat the analysis for the case of maximum available reflected flux ($\phi=0.45$) and for that of maximum shift in radial velocity ($\phi=0.25$) between $F_1$ and $F_2$. Figure~\ref{fig:ica_045}  shows the  first and second extracted components, and their ACF's, as obtained by our computation once FastICA is run with the gaussian contrast function  $G_{2}(u)$ (see  equation~\ref{eq:gfunc}), $SNR=50,000$, and ICASSO with $M=15$ iterations. 
 In both cases, as can be seen already by a visual comparison, ICA is able to disentangle the two components (see (a),(c) and (b),(d) panels, respectively).

The comparison of the peaks of the ACF profiles, shown in (e) and (f) panels, with the no-detection case shows that:
\begin{itemize}
     \item in the case of the maximum available flux both components are retrieved and reconstructed; peak of the ACF profiles for the first component is close to
    $1$, thus the detection can be considered optimal ($D_1 \simeq 100$); the
    detection of the second component is lower 
     ($D_2 \simeq65$), but still the
     albedo trend is clearly noticeable in  panel (c);
    \item when the system is at $\phi=0.25$ the retrieved second component is much more noisy (see panel (d)) and this is reflected by a decrease in the peak of the corresponding ACF profile. Nevertheless, also in this case the detection of the planet signal is achieved, $D_2\simeq20$.
\end{itemize}


Figure~\ref{fig:ica_045_7000} shows results obtained by using a $SNR=6,500$ where the detection of the second component could still be considered achieved ($D_2\simeq3$). As shown in panel~(c), the peak of the ACF profile for the second component (green curve) indeed still slightly exceeds our assumed limit of $3$ for a reliable detection and confirms what was shown also in Figure~\ref{fig:gfunc}. It is worth to highlight that, also in this case of much lower $SNR$, the first component (panel~(a)) is still optimally retrieved ($D_1 \simeq100$). On the other hand no conclusion can be inferred visually for the albedo wavelength dependency (see panel~(b)) without applying some procedure of noise filtering.

In conclusion, our simulation results show that by using ERLICA we can effectively disentangle the individual components of a composite spectrum of an exoplanetary system like 51~Peg. For a composite spectrum with very high $SNR$ ($>6500$) we
demonstrate that we can be  able to detect the planet signal in a system with a flux ratio on the order of $10^{-4}$. We also show that if the noise in the input spectra is on the order or smaller than the planet signal our method is also capable to provide the planet reflected spectrum  i.e. the albedo wavelength dependence without further processing. For lower $SNR$ the estimate of the wavelength dependence of the planet reflected spectrum would require some extra processing to increase its signal to noise ratio.

\section{results and discussion}\label{sec:planet}

\subsection{The case of the binary system RCMa} \label{sec:simulRCMa}

In order to test our method on real data 
we decided to look at the case of eclipsing binary stars which has the advantage of an higher flux ratio than  in the case of an exoplanetary system. It should be noted indeed 
that for our  approach the differences between an exoplanetary (binary) system and  an eclipsing star (binary) system are really minor:
\begin{itemize}
    \item all the theoretical framework for radial velocity computation described in section~\ref{sec:vrad} applies, 
    without losing in generality, to systems composed by two stars (basically, in the  formulas it is enough 
    to replace the subscript \textit{planet} with \textit{star2});
    \item all the photometric variations due to phase dependent reflected light (section~\ref{sec:phase})
    are, on the contrary, not relevant and can be neglected but the two spectra $F_{\rm star1}(\lambda)$ and $F_{\rm star2}(\lambda)$ can still be  considered as two independent components.
\end{itemize} 
The observed mixture $\mathbf{X}$  in this case is composed  by the following two signals:
\begin{eqnarray} \label{eq:s_icaRCMa}
     & F_{1}(\lambda) ~~~ & =  F_{\rm star1}(\lambda) \nonumber \\
     & F_{2}(\lambda,\phi) & =  F_{\rm star1}(\lambda) +  F_{\rm star2}(\lambda,\phi) = F_{\rm star1}(\lambda) +
      F_{star2}\left(\lambda\left[ 1+\frac{RV_{\rm star2}(\phi)-RV_{\rm star1}(\phi) }{c}\right]\right) \nonumber
\end{eqnarray}
where $F_{1}(\lambda)$ is the spectrum of the system in secondary eclipse, $0.45<\phi<0.55$ (i.e. a spectrum taken when the secondary star is hidden behind the primary), and 
$F_{2}(\lambda,\phi)$ is the combined spectrum of the primary and of the secondary obtained at $0.1<|\phi|<0.4$, thus avoiding the spectra taken during the primary eclipse where the problem of limb-darkening \citep[see][]{Winn10} and the
Rossiter-McLaughlin \citep{Rossiter24,McLaughlin24} effect modify the observed $F_{\rm star1}$. We recall that the $F_{\rm star2}$ contribution to $F_2(\lambda,\phi)$ at each phase is Doppler shifted in wavelength because of the  radial velocity of the secondary star with respect to the primary one.
The required statistical independence of the two components hidden in the composite spectra  is guaranteed  because of the   shift in radial velocity of the two stellar spectra and even  reinforced by the fact that, in most astronomical case,  the two stars belong to  different spectral classes and, therefore, have quite different spectra.

Out of the several  eclipsing binary systems described in  literature we decided to study R\,CMa. The eclipsing binary star R\,CMa is a short-period Algol-type system showing an extraordinary small mass ratio between its components. R\,CMa was known for a long time as the system of lowest total mass and as the prototype of a small group of stars called the R\,CMa-type stars, introduced by \citet{1956AnAp...19..298K} and characterized by low mass ratio, overluminosity of the primary, and oversized secondary. 
\citet{2011MNRAS.418.1764B} give a comprehensive overview on the history of the investigations of R\,CMa. They performed a combined photometric, astrometric, and spectroscopic analysis of the R\,CMa system and end up with the stellar parameters given in Table\,\ref{tab:RCMa}. 

\begin{deluxetable*}{ccl}[ht]
\tablecaption{Stellar parameters for R\,CMa system \label{tab:RCMa}}
\tablecolumns{2}
\tablewidth{0pt}
\tablehead{
\colhead{Parameter} &
\colhead{Value} &
\colhead{Reference}
}
\startdata
$M_1$ (M$_\odot$) & $1.67\pm 0.08$ & \citet{2011MNRAS.418.1764B} \\
$M_2$ (M$_\odot$) & $0.22\pm 0.07$ & \citet{2011MNRAS.418.1764B} \\
$q$ (mass ratio) & $0.13\pm 0.05$ & \citet{2011MNRAS.418.1764B} \\
$R_1$ (R$_\odot$) & $1.78\pm 0.03$ & \citet{2011MNRAS.418.1764B} \\
$R_2$ (R$_\odot$) & $1.22\pm 0.07$ & \citet{2011MNRAS.418.1764B} \\
$T_{\rm eff}$ (primary) (K) & $7300$ & \citet{2011MNRAS.418.1764B} \\
$T_{\rm eff}$ (secondary) (K) & $4350$ & \citet{2011MNRAS.418.1764B} \\
$T_{\rm eff}$ (primary) (K) & $7033 \pm 42$ & \citet{Lehmann18} \\
$T_{\rm eff}$ (secondary) (K) & $4350 \pm 100$ & \citet{Lehmann18} \\
\enddata
\end{deluxetable*}

\citet{Lehmann18} used time series of high-resolution spectra and analyze the decomposed spectra of the components together with the radial velocities obtained from decomposed, least-squares deconvolved mean line profiles (LSD profiles, see \citet{1997MNRAS.291..658D}). Their results confirm the values given by \citet{2011MNRAS.418.1764B} for the masses and radii, and also for the $T_{\rm eff}$ of the secondary component, whereas the $T_{\rm eff}$ derived for the primary component is by 300\,K lower (see Table\,\ref{tab:RCMa}). Authors did not find evidence of the presence of a third body in the system, as supposed by \citet{1984BASI...12..182R} or \citet{2002AJ....123.2033R}.

We have chosen R\,CMa as our test star because of its  luminosity ratio,$\frac{F_{\rm star2}}{F_{\rm star}} \simeq 0.04$ in the visible,   and because we can compare the results of our  method with those obtained by \citet{Lehmann18}. The latter authors used high-resolution spectra of R\,CMa obtained with the
HERMES spectrograph \citep{Raskin2011}. The spectra were reduced using the standard HERMES
pipeline and, subsequently,  normalized to the local
continuum. Then the Fourier transformation-based KOREL program \citep{1995A&AS..114..393H,Hadrava2006} was used
to disentangle the observed composite spectra. In fact, from a time series of spectra, the program delivers the decomposed spectra of the components, normalized to the common continuum of both stars, together with the optimum orbital elements, assuming pure Keplerian orbits. The spectra of the components, resulting from observations in all out-of-eclipse phases, were renormalized to the individual continua by help of the wavelength dependent continuum flux ratio which was  derived from spectrum analysis.

\begin{figure}[ht]
\plotone{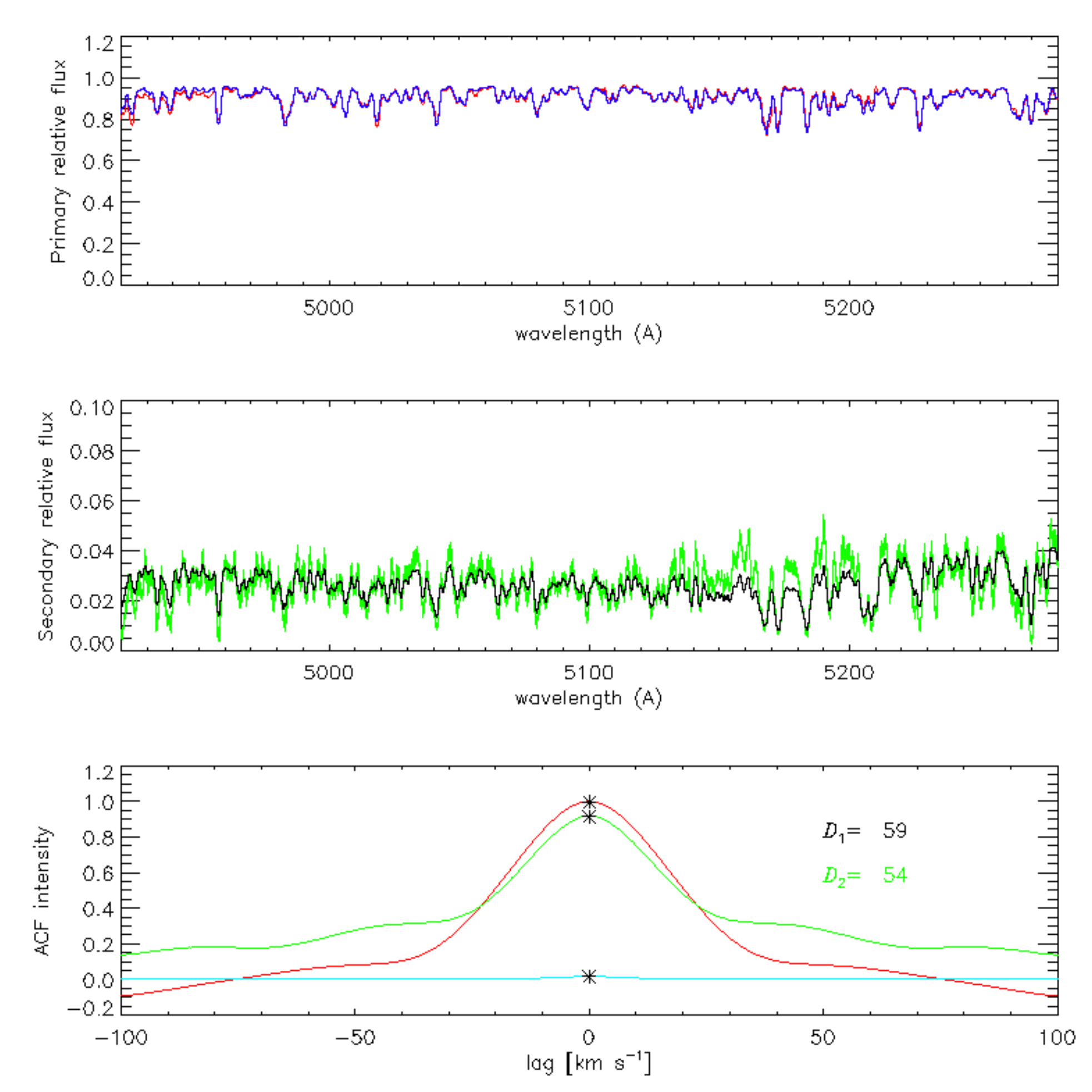}
\caption{ICA results for the RCMa system: comparison between the extracted first component (red) and the synthetic spectrum of the primary star (blue) - upper panel; comparison between the extracted second component (green) and the synthetic spectrum of the secondary star (black) - middle panel;
ACF profiles of averaged first (red) and second (green) components compared with the no-detection case (light blue) - lower panel; \label{fig:RCMa}}
\end{figure}

In our analysis we used the same HERMES normalized spectra used by \citet{Lehmann18} and we applied our method to  each \textit{j} possible pair of $F_{1}(\lambda)$  and $F_{2}(\lambda,\phi)$  spectra. Thus, after every ERLICA run, i.e for each \textit{j} pair of spectra, we checked the detection significance of the derived components, $S_1^j$ and $S_2^j$. Then we properly averaged them to build $\langle S_1\rangle$ and $\langle S_2\rangle$, i.e. our final estimates of $F_{\rm star1}$ and $F_{\rm star2}$, respectively. In making the average of the second component estimates we used the $RV_{\rm star2,star1}$ from \citet{Lehmann18} to put all the individual $S_2^j$ in the reference frame where $RV_{\rm star2}=0$. The results are shown in Figure~\ref{fig:RCMa}: in the top panel we plot $\langle S_1\rangle$ compared with the synthetic spectrum of the primary star computed using its atmospheric parameters given in \citep[][Table~1]{Lehmann18}, in the middle panel $\langle S_2\rangle$ is compared with the corresponding synthetic spectrum, and in the bottom panel we show  $\langle S_1\rangle$ an $\langle S_2\rangle$  ACFs. To evaluate the detection significance we show in the bottom panel also the ACF of the \textit{false} $\langle S_2\rangle$ we obtained by using  \textit{k} pairs built with  two  spectra both taken during the secondary eclipse.
As can be seen the $\langle S_1\rangle$ and $\langle S_2\rangle$ are in very good agreement with the corresponding synthetic spectra and  their detection significance is $D_1\simeq59$ and $D_2=\simeq54$, respectively.

\begin{figure}[ht]
\plotone{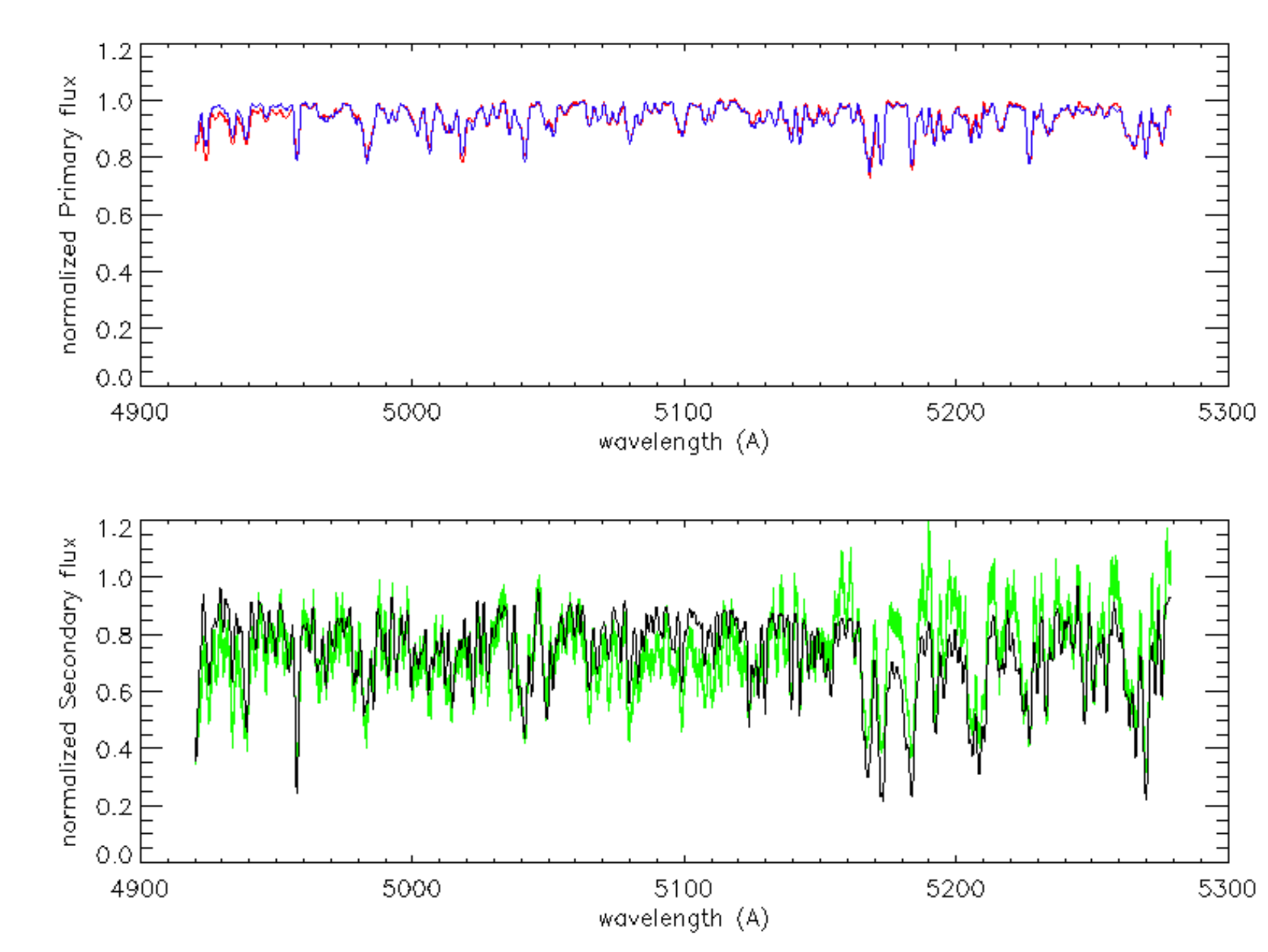}
\caption{Comparison between the extracted spectrum of the Primary (top panel) and Secondary (bottom panel) of R\,CMa as obtained in this paper (red and green) and those obtained by \citet{Lehmann18} (blue and black)  \label{fig:Lehmann} using KOREL.}
\end{figure}

 A comparison of our results and those obtained by \citet{Lehmann18} using KOREL is shown in Figure \ref{fig:Lehmann}. As can be seen there is a very good agreement for the Primary spectrum (with an $rms$ values of 0.01) and a satisfactory agreement for the Secondary spectrum ($rms = 0.10$).
 The obtained $rms$ values are on the same order of those between the derived spectra and the corresponding synthetic ones. We recall that the validity of the synthetic spectra  is limited by the physics included in the corresponding models and by the  different program codes used to calculate them. Figure~\ref{fig:Lehmann} demonstrates that our method provides estimates of the disentangled spectra as reliable of those obtainable by well proofed programs used in binary star analysis.
 
 In conclusion, combining the results of our simulations (see Section~\ref{sec:simul}) and those obtained  in the case of R\,CMa, we can say that the validity of our method is assessed.
 
\subsection{The case of 51\,Peg}

As shown in section~\ref{sec:gfunc}, to reliably disentangle individual components of binary systems using our ICA-based method, it is mandatory to have at disposal spectra  with very high $SNR$ . This is particularly demanding in the case of exoplanetary systems which are characterized by a very low flux ratio  
(of the order of $1 \times 10^{-4}$ or less); simulations (see section~\ref{subsec:sim_res}) show that, for a typical case, an
$SNR>5000$ is required. To reach such an $SNR$ is probably outside the possibility of the current instrumentation and certainly not available in currently public available data.

A  way to mitigate such limitation with the aim to successfully apply our method also on such demanding systems 
is to try to increase somehow the $SNR$. This can be achieved, for example, by:
\begin{itemize}
    \item using averages of multiple spectra instead of single ones for the $F_1(\lambda)$ and $F_2(\lambda)$ in equation \ref{eq:x_ica} with the constrain that they should be taken at exactly the same $\phi$ values;
   \item using more than one pair of $F_1(\lambda)$ and $F_2(\lambda)$ and, then, averaging the retrieved components as done in Section~\ref{sec:simulRCMa}.
  
  \end{itemize}
  Based on these considerations we decided to test our method on the real data of an exoplanetary system and in particular  we decided to use 91 HARPS spectra of the system 51~Peg~$+$~51~Peg\,b already used by \citet{Martins15} and by \citet{Borra18}. The spectra were re-reduced by using HARPS DRS~3.4 and kindly made available to us by J.H.C. Martins.
  \citet{Martins15},  using the CCF method,  and \citet{Borra18},  using the ACF method,  detected from the analysis of these spectra the signal of 51~Peg\,b with a detection significance of 3.70~$\sigma_{\rm noise}$ and 5.52~$\sigma_{\rm noise}$, respectively.

\begin{figure}[ht]
\plotone{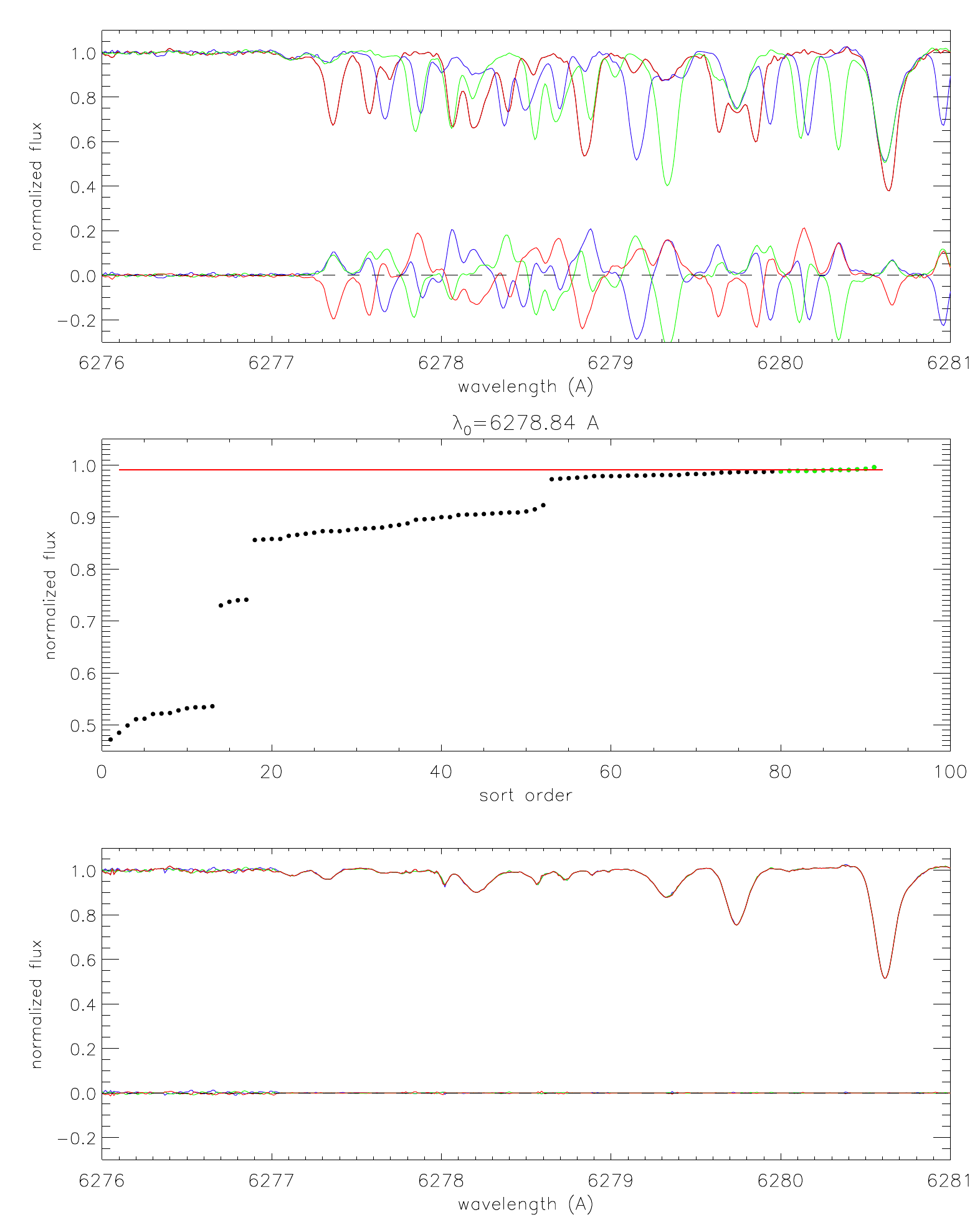}
\caption{An example of the adopted telluric line correction: three spectra  of 51~Peg obtained at different heliocentric velocities (red, blue and green) and the differences from
their mean, before (upper panel) and after removing the telluric lines (bottom panel);
central panel shows the plot of the sorted fluxes used to select ``uncontaminated'' spectra (green points) and the estimated uncontaminated flux value (red line). See text for details.\label{fig:telluriche}}
\end{figure}

  We limited our analysis to the wavelength region $5400$\,\AA\,$<\lambda\,<6800$\,\AA~ where the  spectra
  have  $SNR>200$.
  Unfortunately this wavelength range is affected in several regions by the presence of telluric lines and  remove them is not an easy task \citep[see discussion in][]{Smette2015}.
  The method we applied is based on the fact that the HARPS spectra were  obtained on different Julian days and, thus, they are affected by different heliocentric velocities. Therefore, after correcting all the spectra for the proper heliocentric velocity to put them in the wavelength laboratory rest frame,  and after normalization, each spectrum shows the contamination of telluric lines at different wavelength positions (see upper plot in the upper panel of Figure\,\ref{fig:telluriche}). 
   These wavelength positions can be easily identified by their anomalously large standard deviations with respect to the mean spectrum. Then,  by sorting at  each individual position the normalized flux of all the spectra, we separated the spectra which, in that specific point, have the higher signals from the others. The former are those less affected by  telluric lines and their mean intensity value was  used  as an estimate of the telluric corrected flux value. Eventually, this value was adopted to substitute, at the corresponding  wavelength point, the signal in the latter ones.
   
   An example of the correction procedure in one of the region affected by telluric lines is shown in Figure~\ref{fig:telluriche} where:.
   \begin{itemize}
   \item the upper panel shows three spectra  and the differences from their mean before our telluric line correction;
   \item the central panel shows the normalized sorted flux values at one of the $\lambda$ in the region, $\lambda_0=6278.84$\,\AA. This plot is used to select the "uncontaminated" spectra (green dots) and to compute an  estimate of the telluric corrected flux value at $\lambda_0$(red line). The number of ``uncontaminated'' spectra, 12, is a good compromise obtained by visual inspecting several plots like that one shown;
   \item the bottom panel shows the same three spectra of the left panel after the correction with, at the bottom, the new differences from their mean.
   \end{itemize}
  As can be seen the adopted method works quite well even if some  residual contamination (on the order of few percents in normalized fluxes) still remains in very critical regions (see lower plot in the bottom  panel of Figure\,\ref{fig:telluriche}.
  
 To perform our ERLICA analysis the 91 HARPS spectra were Doppler shifted for the $RV_{\rm star}$ values and divided in two groups: 
  \begin{itemize}
      \item Group~1: it contains the 20 spectra that are near the inferior conjunction; they can be used, individually, as $F_1(\lambda)$ in equation \ref{eq:x_ica};
      \item Group~2: it contains the  71 spectra that are supposed to contain both $F_{\rm star}$ and $F_{\rm planet}$ since they have been observed out of the inferior conjunction; they can be used as $F_2(\lambda)$ in equation \ref{eq:x_ica}.
  \end{itemize} 
  Now we are able to follow  the steps outlined in Section~\ref{sec:simul1}. In particular:
\begin{enumerate}\renewcommand{\labelenumi}{(\roman{enumi})}
\setlength{\itemindent}{0.25in}
\setcounter{enumi}{2}
\item create the observed mixture $\mathbf{X}$ using for $F_1(\lambda)$ a spectrum pertaining to group~1 and for $F_2(\lambda)$ a spectrum of group~2 and apply ERLICA;
\item fix signs and order of the extracted components  $S_{\rm star}$ and $S_{\rm planet}$ to obtain estimates of $F_{\rm star}$ and $F_{\rm planet}$;
\item compute ACF of each component after re-sampling  at steps of constant velocities;
\item compute detection significance $D_{\rm star}$ and $D_{\rm planet}$ values.
\end{enumerate}
 With the aim of increasing the $SNR$ we paired each of the   20 $F_1(\lambda)$ spectra individually with  all the 71 spectra of group~2.  Eventually we were able to apply ERLICA on $20 \times 71=1420$ 
  mixtures $X_j$ and for
  each \textit{j} pair of $F_1(\lambda)$ and $F_2(\lambda)$ spectra, we obtained  the $S_{\rm star}^j$ and $S_{\rm planet}^j$ estimates of $F_{\rm star}$ and $F_{\rm planet}$. Moreover, to compute $D_{\rm star}$ and $D_{\rm planet}$, we applied ERLICA  on the 190 $k$ pairs built using as $F_2(\lambda)$ a spectrum of group~1 in order to check our $F_{\rm planet}$ estimate. In fact, in these cases we shouldn't  detect any signature from the planet and the analysis of the $k$ obtained second ICA components  can  give estimates of  possible ``fake/spurious'' $S_{\rm planet}$ introduced by ERLICA.
  \subsubsection{Results}
\begin{figure}[ht]
\plotone{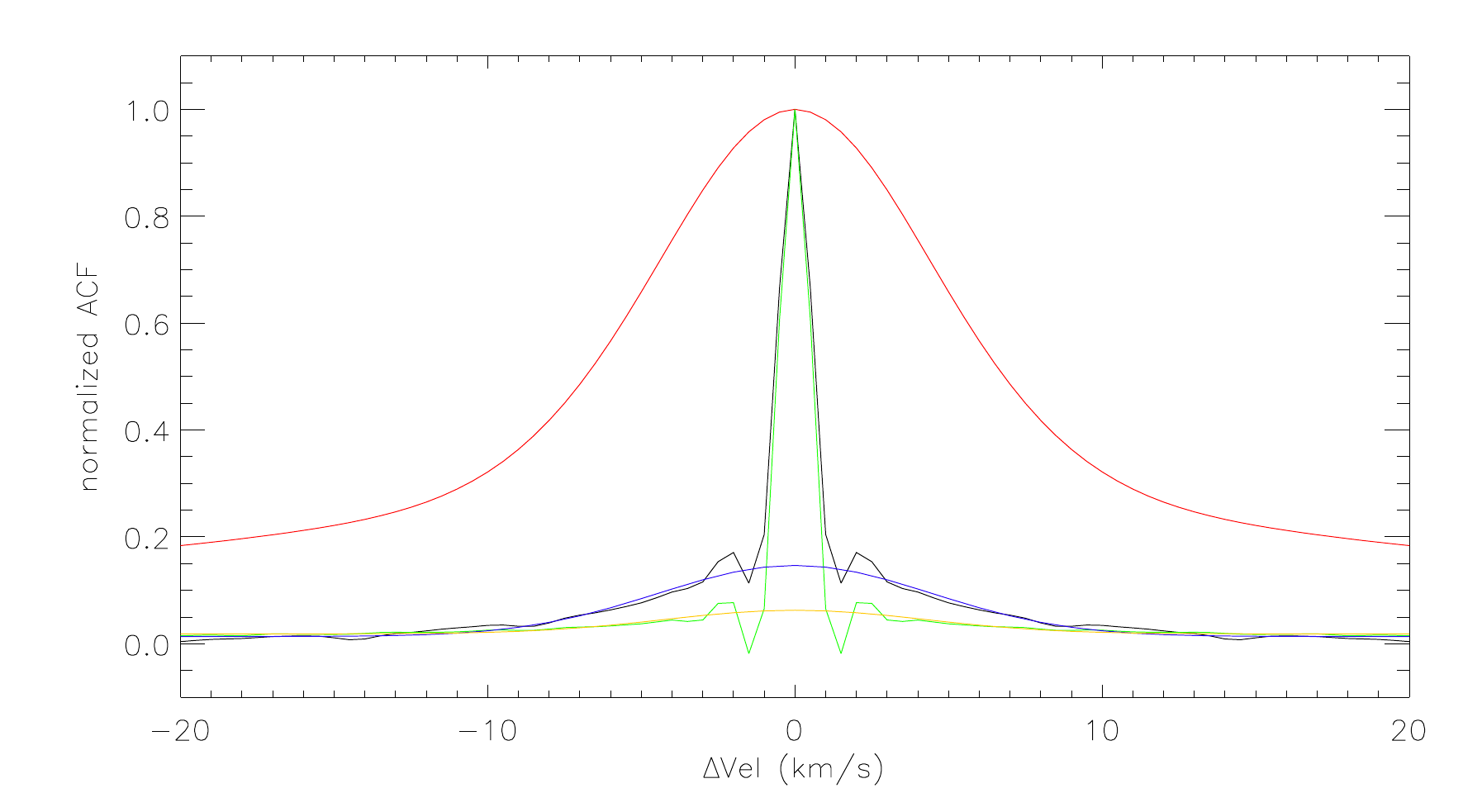}
\caption{
ACF profiles of average first (red) and second (black) components compared with the ACF of the ``fake'' second component (green) The blue and yellow lines are the fit of the black and green ACFs after removing the central points (see text). \label{fig:51Peg_ACFs}}
\end{figure}

  Figure~\ref{fig:51Peg_ACFs} shows the average auto-correlation functions of the $S_{\rm star}^j$, 
  $S_{\rm planet}^j$, and ``fake'' $S_{\rm planet}^k$.
As can be seen   the mean ACF of $S_{\rm star}^j$ shows that our procedure derives a well defined
  $S_{\rm star}$ component which provides a very accurate estimate of $F_{\rm star}$.
  The identification of $S_{\rm star}$ with $F_{\rm star}$ only is confirmed by the absence of any planetary signal if we apply to the $S_{\rm star}^j$ ACFs the same analysis used by \cite{Borra18} to detect the signal from 51Peg\,b (see their Figure~7)
  
  As far as the average ACF of the $S_{\rm planet}^j$ is concerned the computation of its detection significance, as described in Section~\ref{subseq:extraction} requires to remove the central peak. Actually both the  ACFs of the $S_{\rm planet}^j$ and of the ``fake''  $S_{\rm planet}^k$ show  central peaks which extend from -2 to +2\, km\,s$^{-1}$ suggesting the presence of some correlated noise. Thus we removed this velocity interval before computing the gaussian fits shown by the blue and yellow lines in Figure\,\ref{fig:51Peg_ACFs}. The obtained $D_{\rm planet}=2.3$ value is below our adopted detection threshold ($D=3$) and this result is qualitatively in agreement with those obtained from  our simulation (see Figure~\ref{fig:gfunc}). In fact, even if we assume that our $F_1(\lambda)$,$F_2(\lambda)$ pairs are uncorrelated (which is quite unlike) the SNR of our spectra, $\sim 200$, should be multiply by the square root of the number of pairs, $\sqrt{1420}$, leading to a value
  of $SNR\sim 7500$ which is slightly above the $SNR$ limit from Figure\,\ref{fig:gfunc}. 
  Thus, taking into account the not complete independence of our $F_1(\lambda)$,$F_2(\lambda)$ pairs and the different values of the $SNR$ of the individual HARPS spectra, we can conclude that a low detection significance was expected. Furthermore the spectra coverage of the planetary orbit is neither complete nor homogeneous making the used HARPS data not an optimal set  for our method (in  particular we have very few spectra taken close to the superior conjunction or at  phases near the maxima of  $abs(RV_{\rm planet,star})$).
  In any case, it is worthwhile to notice that if we had used an estimator of detection similar to those used in CCF and ACF approaches (for example  the ratio between the peak of the fitted ACF  and the standard deviation of the ACF pixel intensity for $abs(\Delta\,{\rm vel}) > 10$\,km/s) we would have obtained $D_{\rm noise}= 14.3\, \sigma_{\rm noise}$, i.e. a value larger than those obtained by \citet{Martins15} or by \citet{Borra18}.

 Even if the detection significance, $D_{\rm planet}$, of the second component is quite low we tried to derive its wavelength dependence in order to understand its nature, i.e. if it contains the $F_{\rm planet}$ signature or if it is mainly due to systematics. To do that the individual $S_{\rm planet}^j$ estimates were averaged after applying the proper Doppler shifts to put them in a reference system where $RV_{\rm planet}=0$.
 
\begin{figure}[ht]
\plotone{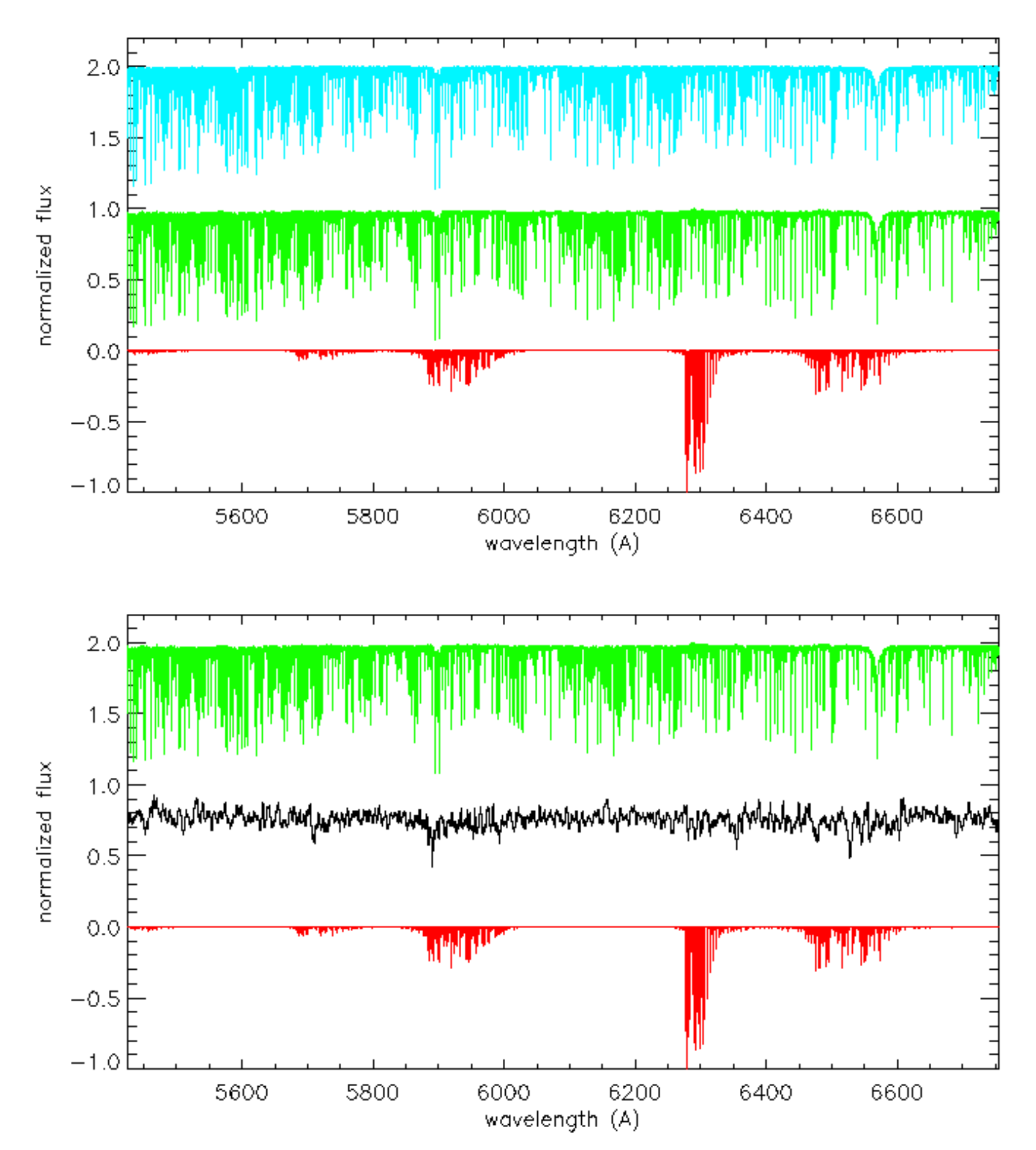}
\caption{ICA results for the 51~Peg system in the range $5400\,$\AA~$<\lambda<6800\,$\AA. Upper panel: comparison among the extracted first component (green), the synthetic spectrum of the star (light blue), and a scaled ($2\times$) SKYCALC spectrum (red); lower panel: extracted and smoothed second component (black),  extracted first component (green), and  SKYCALC spectrum (red). The spectra are vertically shifted to increase the readability.
\label{fig:51Peg5400}}
\end{figure}

 Figure~\ref{fig:51Peg5400} illustrates the final results:
 \begin{itemize}
 \item in the top panel we plot in green the average estimate of $S_{\rm star}^j$, the synthetic spectrum of 51~Peg used in Section\,\ref{sec:simul1} in light blue, and, to show the regions affected by telluric line contamination we may have not corrected perfectly,  a scaled ($2 \times$) telluric spectrum computed with SKYCALC\footnote{https://www.eso.org/observing/etc/bin/gen/form?INS.MODE=swspectr+INS.NAME=SKYCALC} in red;
 \item  the bottom panel contains the average estimate of $S_{\rm planet}^j$ in black, the average estimate of $S_{\rm star}^j$ in green, and the SKYCALC spectrum in red. The black curve was smoothed by using a 200~points running average to reduce the noise and shows some relatively broad features.
 \end{itemize}
 
\begin{figure}[ht]
\plotone{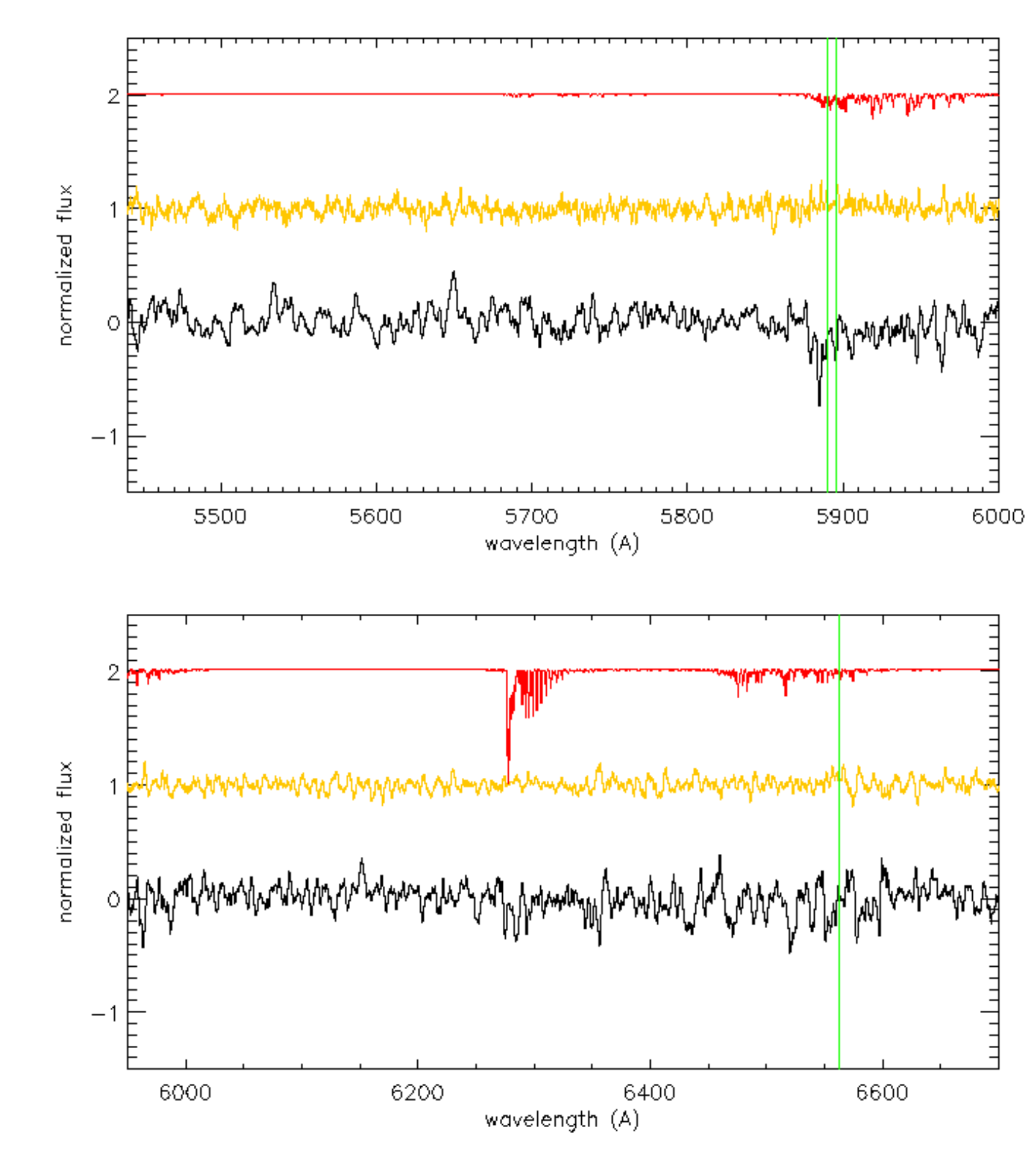}
\caption{Average estimates of $S_{\rm planet}^j$ (black, possible $F_{\rm planet}$)  and $S_{\rm planet}^k$  (yellow, inferior conjunction, where the reflected $F_{\rm planet}$ signal cannot be present), and the SKYCALC telluric spectrum (red). The whole wavelength range was divided in two parts (upper and lower panels) and the spectra are vertically shifted to increase the readability. The three vertical lines show the laboratory positions of the Na\,D doublet lines and of H$\alpha$.
\label{fig:51Peg5400b}}
\end{figure}

 Whether the average estimate of $S_{\rm planet}^j$ is the real reflected signal from 51~Peg\,b is however questionable. In an attempt to answer to this question we compare in Figure~\ref{fig:51Peg5400b} the average estimates of $S_{\rm planet}^j$ (black) with the ``fake'' $S_{\rm planet}^k$  (yellow), derived from spectra in the inferior conjunction where the reflected $F_{\rm planet}$ signature cannot be present. 
 Due to the centering and whitening of the input data and of the ambiguities intrinsic in ICA described in 2.3 we didn't try to recover the absolute value of the black and yellow spectra and, therefore, both have zero mean value (in Figure~\ref{fig:51Peg5400b}  the spectra were vertically shifted to increase the readability).
 
 As can be seen, there is a very low similarity between the star spectrum and the extracted and smoothed second component (see lower panel of Figure~\ref{fig:51Peg5400}). In particular the strongest lines in the 51~Peg spectrum, i.e. the Na\,D doublet and H$\alpha$ are not present in the black curves of Figure~\ref{fig:51Peg5400b}. This is in contradiction with the expected results (see Figure~\ref{fig:ica_045}) where the reflected signal should consist of the stellar spectrum modulated by the planetary albedo.
 On the other hand, the presence of more evident features and the corresponding larger standard deviations of the black curves with respect to the yellow ones in Figure~\ref{fig:51Peg5400b} seems to suggest that we indeed detect some signal from 51~Peg\,b.
 If this is true a possible explanation for the absence of the stellar lines in our average second component could be the presence of clouds in the atmosphere of 51~Peg\,b which may smooth and flatten the planet reflected spectrum \citep[see discussion, for example in][]{GAO2017}. 
 Another possibility is that the ``detected'' features are the remnant of not completely removed telluric lines since the strongest of them fall in the critical regions where the SKYCALC  spectrum shows most of the telluric lines (see Figure~\ref{fig:51Peg5400b}).
 
 In conclusion, we do not have a sound final answer about the nature of the features in the  average extracted second component. The possible detection of the reflected spectrum of 51~Peg\,b to be confirmed  would require to repeat our analysis using new ``ad-hoc'' obtained input data, i.e. at higher $SNR$, covering the whole range of orbital phases, in particular both conjunctions, with more than one spectrum at each phase, and with auxiliary spectra to be used for accurately removing the telluric contamination. 
  
\section{Conclusions} \label{sec:conclusion}
In this paper we presented a new method based on applying the Independent Component Analysis technique
for extracting the reflected planetary spectral signature from a series of composite spectra of a binary exoplanetary system.
The main advantages, compared to the commonly adopted techniques like CCF ad ACF, are that the extraction is \textit{``blind''} i.e. it does not require any \textit{a priori} knowledge of the underlying signals 
and that the method allows not only to detect the presence of a planet contribution, but also to estimate its wavelength dependence.

To show and quantify the validity and effectiveness of the proposed approach, ERLICA, we applied it first on  simulated data of an exoplanetary system with physical characteristics similar 
to 51~Peg~+~51~Peg\,b. In section~\ref{subseq:extraction} we introduced a quantitative estimator $D$ of the detection significance  to asses quantitatively the ERLICA disentangling capability. The results of the simulation showed that our methods provides, in any case,
an accurate estimate of the stellar spectrum and, when the noise in the input spectra is on the order or smaller than the planet signal, also the planetary albedo wavelength dependence.

Then we analyzed successfully
the real case of an eclipsing binary star, the R\,CMa system, 
taking advantage of the fact that a binary star system  could be considered physically similar to an exoplanetary system, but with  a much higher flux ratio of the two components. The results for the spectra of the primary and secondary star showed that our method can be considered as a valid, and somewhat easier, alternative to well established codes for analysing binary stars like e.g. KOREL, FDbinary \citep{Ilijic04} or Spectangular \citep{Sablowski17}.

Eventually we applied the method  on real 51~Peg~+~51~Peg\,b data. 
Also in this case we showed that our method is capable to extract the stellar spectrum very effectively. As far as the detection of the planetary signal is concerned we obtained a quite low detection significance, $D_{\rm planet}=2.3$, even if it is worthwhile to point out that if we had used an estimator similar to those used by \citet{Borra18} and \citet{Martins15} we would have obtained a value which is larger than
those obtained with the CCF or ACF methods. 

Unfortunately, our attempts to analyze the wavelength dependence of the ``possible'' reflected spectrum of 51~Peg\,b to confirm its nature gave not conclusive results due to insufficient $SNR$ and to the absence of auxiliary data needed to accurately remove the telluric contamination.  Therefore we could not definitively proof that we were able to derive the  reflected  spectrum of 51~Peg\,b.

In conclusion we can say that the proposed ERLICA approach could be considered, at least in perspective, a powerful tool for studying and characterizing exoplanetary systems. In fact, we want to point out that  this method will benefit significantly from  the availability, in the near future, of new ``ad-hoc''  scheduled observations obtained with the just coming into operation state-of-the-art instrumentation for exoplanetary research like ESO/VLT ESPRESSO \citep{espresso}, 
as well as with the foreseen instrumentation for the forthcoming 30m class telescopes 
(like the High Resolution Spectrograph, HIRES, for the  ESO ELT \citep{hires}). 
In fact, these new instruments, due to their higher efficiency and telescope larger effective area, would allow to obtain spectra with much higher $SNR$ with the same exposure time than those used in this paper ($SNR_{\rm HARPS} \simeq 200$, $SNR_{\rm ESPRESSO} \simeq 500$, $SNR_{\rm HIRES} > 9000$) as derived from the corresponding Exposure Time 
Calculators\textsuperscript{  }\footnote{https://www.eso.org/observing/etc/}\textsuperscript{, }\footnote{https://www.arcetri.astro.it/$\sim$hires/etc.html} thus allowing to fully exploit the ERLICA capabilities.

\acknowledgments
  HL acknowledges support by the DFG grant LE1102/3-1.
 We want to thank J.H.C. Martins for providing us the up-dated reduced data of 51~Peg.
 P.D.M. thanks the ESPRESSO Science Working Group 2 team and in particular M. R. Zapatero Osorio, N. C. Santos, F. Pepe, C. Lovis and D. Ehrenreich, for hints and fruitful discussions.
 We thanks also several graduate students, in particular R. Bevilacqua and  P. Menia, of the Universit\`a degli Studi di Trieste (Italy) who helped us, in the framework of their curricular internships, in setting the simulation used in this paper and in the analysis of RCMa.
 
 \facilities{ESO:HARPS}
 \software{FastICA \citep[v2.5;][]{Hyvarinen99}, 
           ICASSO \citep[v1.21;][]{Himberg04}}

\bibliographystyle{aasjournal}
\bibliography{mybiblio}

\begin{thebibliography}{}
\expandafter\ifx\csname natexlab\endcsname\relax\def\natexlab#1{#1}\fi
\providecommand{\url}[1]{\href{#1}{#1}}

\bibitem[{{Belouchrani} {et~al.}(1997){Belouchrani}, {Abed-Meraim}, {Cardoso},
  \& {Moulines}}]{Belouchrani97}
{Belouchrani}, A., {Abed-Meraim}, K., {Cardoso}, J.~., \& {Moulines}, E. 1997,
  IEEE Transactions on Signal Processing, 45, 434

\bibitem[{{Borra} \& {Deschatelets}(2018)}]{Borra18}
{Borra}, E.~F., \& {Deschatelets}, D. 2018, \mnras, 481, 4841

\bibitem[{{Budding} \& {Butland}(2011)}]{2011MNRAS.418.1764B}
{Budding}, E., \& {Butland}, R. 2011, \mnras, 418, 1764

\bibitem[{{Cardoso} \& {Souloumiac}(1993)}]{Cardoso93}
{Cardoso}, J.~F., \& {Souloumiac}, A. 1993, IEEE Proceedings F (Radar and
  Signal Processing, 140, 462

\bibitem[{{Charbonneau} {et~al.}(1999){Charbonneau}, {Noyes}, {Korzennik},
  {Nisenson}, {Jha}, {Vogt}, \& {Kibrick}}]{Charbonneau99}
{Charbonneau}, D., {Noyes}, R.~W., {Korzennik}, S.~G., {et~al.} 1999, \apjl,
  522, L145

\bibitem[{{Collier Cameron} {et~al.}(1999){Collier Cameron}, {Horne}, {Penny},
  \& {James}}]{CollierCameron99}
{Collier Cameron}, A., {Horne}, K., {Penny}, A., \& {James}, D. 1999, \nat,
  402, 751

\bibitem[{{Donati} {et~al.}(1997){Donati}, {Semel}, {Carter}, {Rees}, \&
  {Collier Cameron}}]{1997MNRAS.291..658D}
{Donati}, J.-F., {Semel}, M., {Carter}, B.~D., {Rees}, D.~E., \& {Collier
  Cameron}, A. 1997, \mnras, 291, 658

\bibitem[{{Fuhrmann} {et~al.}(1997){Fuhrmann}, {Pfeiffer}, \&
  {Bernkopf}}]{Fuhrmann97}
{Fuhrmann}, K., {Pfeiffer}, M.~J., \& {Bernkopf}, J. 1997, \aap, 326, 1081

\bibitem[{{Gao} {et~al.}(2017){Gao}, {Marley}, {Zahnle}, {Robinson}, \&
  {Lewis}}]{GAO2017}
{Gao}, P., {Marley}, M.~S., {Zahnle}, K., {Robinson}, T.~D., \& {Lewis}, N.~K.
  2017, \aj, 153, 139

\bibitem[{{Gray} \& {Corbally}(1994)}]{Gray94}
{Gray}, R.~O., \& {Corbally}, C.~J. 1994, \aj, 107, 742

\bibitem[{{Hadrava}(1995)}]{1995A&AS..114..393H}
{Hadrava}, P. 1995, \aaps, 114, 393

\bibitem[{{Hadrava}(2006)}]{Hadrava2006}
---. 2006, \apss, 304, 337

\bibitem[{{Hadrava}(2016)}]{Hadrava16}
{Hadrava}, P. 2016, in Astrophysics and Space Science Library, Vol. 439,
  Astronomy at High Angular Resolution, ed. H.~M.~J. {Boffin}, G.~{Hussain},
  J.-P. {Berger}, \& L.~{Schmidtobreick}, 113

\bibitem[{Himberg {et~al.}(2004)Himberg, Hyvärinen, \& Esposito}]{Himberg04}
Himberg, J., Hyvärinen, A., \& Esposito, F. 2004, NeuroImage, 22, 1214 .
\newblock
  \url{http://www.sciencedirect.com/science/article/pii/S1053811904001661}

\bibitem[{{Hyvarinen}(1999)}]{Hyvarinen99}
{Hyvarinen}, A. 1999, IEEE Transactions on Neural Networks, 10, 626

\bibitem[{{Hyv{\"a}rinen} {et~al.}(2001){Hyv{\"a}rinen}, {Karhunen}, \&
  {Oja}}]{Hyvarinen01}
{Hyv{\"a}rinen}, A., {Karhunen}, J., \& {Oja}, E. 2001, {Independent Component
  Analysis} ({Jonh Wiley \& Sons, Inc}), 481

\bibitem[{{Ilijic} {et~al.}(2004){Ilijic}, {Hensberge}, {Pavlovski}, \&
  {Freyhammer}}]{Ilijic04}
{Ilijic}, S., {Hensberge}, H., {Pavlovski}, K., \& {Freyhammer}, L.~M. 2004, in
  Astronomical Society of the Pacific Conference Series, Vol. 318,
  Spectroscopically and Spatially Resolving the Components of the Close Binary
  Stars, ed. R.~W. {Hilditch}, H.~{Hensberge}, \& K.~{Pavlovski}, 111--113

\bibitem[{{Karkoschka}(1998)}]{Karkoschka98}
{Karkoschka}, E. 1998, \icarus, 133, 134

\bibitem[{{Kopal}(1956)}]{1956AnAp...19..298K}
{Kopal}, Z. 1956, Annales d'Astrophysique, 19, 298

\bibitem[{{Kurucz}(2005)}]{Kurucz05}
{Kurucz}, R.~L. 2005, Memorie della Societa Astronomica Italiana Supplementi,
  8, 14

\bibitem[{{Langford} {et~al.}(2011){Langford}, {Wyithe}, {Turner}, {Jenkins},
  {Narita}, {Liu}, {Suto}, \& {Yamada}}]{Langford11}
{Langford}, S.~V., {Wyithe}, J.~S.~B., {Turner}, E.~L., {et~al.} 2011, \mnras,
  415, 673

\bibitem[{{Lehmann} {et~al.}(2018){Lehmann}, {Tsymbal}, {Pertermann},
  {Tkachenko}, {Mkrtichian}, \& {A-thano}}]{Lehmann18}
{Lehmann}, H., {Tsymbal}, V., {Pertermann}, F., {et~al.} 2018, \aap, 615, A131

\bibitem[{{Lindegren} \& {Dravins}(2003)}]{Lindegren03}
{Lindegren}, L., \& {Dravins}, D. 2003, \aap, 401, 1185

\bibitem[{{Madden} \& {Kaltenegger}(2018)}]{Madden18}
{Madden}, J.~H., \& {Kaltenegger}, L. 2018, Astrobiology, 18, 1559

\bibitem[{{Marconi} {et~al.}(2018){Marconi}, {Allende Prieto}, {Amado},
  {Amate}, {Augusto}, {Becerril}, {Bezawada}, {Boisse}, {Bouchy}, {Cabral},
  {Chazelas}, {Cirami}, {Coretti}, {Cristiani}, {Cupani}, {de Castro Le{\~a}o},
  {de Medeiros}, {de Souza}, {Di Marcantonio}, {Di Varano}, {D'Odorico},
  {Drass}, {Figueira}, {Fragoso}, {Fynbo}, {Genoni}, {Gonz{\'a}lez
  Hern{\'a}ndez}, {Haehnelt}, {Hughes}, {Huke}, {Kjeldsen}, {Korn}, {Landoni},
  {Liske}, {Lovis}, {Maiolino}, {Marquart}, {Martins}, {Mason}, {Monteiro},
  {Morris}, {Murray}, {Niedzielski}, {Oliva}, {Origlia}, {Pall{\'e}},
  {Parr-Burman}, {Parro}, {Pepe}, {Piskunov}, {Rasilla}, {Rees}, {Rebolo},
  {Riva}, {Rousseau}, {Sanna}, {Santos}, {Shen}, {Sortino}, {Sosnowska},
  {Sousa}, {Stempels}, {Strassmeier}, {Tenegi}, {Tozzi}, {Udry}, {Valenziano},
  {Vanzi}, {Weber}, {Woche}, {Xompero}, \& {Zackrisson}}]{hires}
{Marconi}, A., {Allende Prieto}, C., {Amado}, P.~J., {et~al.} 2018, in Society
  of Photo-Optical Instrumentation Engineers (SPIE) Conference Series, Vol.
  10702, Ground-based and Airborne Instrumentation for Astronomy VII, 107021Y

\bibitem[{{Marley} {et~al.}(2013){Marley}, {Ackerman}, {Cuzzi}, \&
  {Kitzmann}}]{Marley13}
{Marley}, M.~S., {Ackerman}, A.~S., {Cuzzi}, J.~N., \& {Kitzmann}, D. 2013,
  {Clouds and Hazes in Exoplanet Atmospheres} ({Mackwell}, Stephen J. and
  {Simon-Miller}, Amy A. and {Harder}, Jerald W. and {Bullock}, Mark A.), 367

\bibitem[{{Marley} \& {Robinson}(2015)}]{Marley15}
{Marley}, M.~S., \& {Robinson}, T.~D. 2015, \araa, 53, 279

\bibitem[{{Martins} {et~al.}(2013){Martins}, {Figueira}, {Santos}, \&
  {Lovis}}]{Martins13}
{Martins}, J.~H.~C., {Figueira}, P., {Santos}, N.~C., \& {Lovis}, C. 2013,
  \mnras, 436, 1215

\bibitem[{{Martins} {et~al.}(2018){Martins}, {Figueira}, {Santos}, {Melo},
  {Mu{\~n}oz}, {Faria}, {Pepe}, \& {Lovis}}]{Martins18}
{Martins}, J.~H.~C., {Figueira}, P., {Santos}, N.~C., {et~al.} 2018, \mnras,
  478, 5240

\bibitem[{{Martins} {et~al.}(2015){Martins}, {Santos}, {Figueira}, {Faria},
  {Montalto}, {Boisse}, {Ehrenreich}, {Lovis}, {Mayor}, {Melo}, {Pepe},
  {Sousa}, {Udry}, \& {Cunha}}]{Martins15}
{Martins}, J.~H.~C., {Santos}, N.~C., {Figueira}, P., {et~al.} 2015, \aap, 576,
  A134

\bibitem[{{McLaughlin}(1924)}]{McLaughlin24}
{McLaughlin}, D.~B. 1924, \apj, 60, doi:10.1086/142826

\bibitem[{{M{\'e}gevand} {et~al.}(2014){M{\'e}gevand}, {Zerbi}, {Di
  Marcantonio}, {Cabral}, {Riva}, {Abreu}, {Pepe}, {Cristiani}, {Rebolo Lopez},
  {Santos}, {Dekker}, {Aliverti}, {Allende}, {Amate}, {Avila}, {Baldini},
  {Bandy}, {Bristow}, {Broeg}, {Cirami}, {Coelho}, {Conconi}, {Coretti},
  {Cupani}, {D'Odorico}, {De Caprio}, {Delabre}, {Dorn}, {Figueira}, {Fragoso},
  {Galeotta}, {Genolet}, {Gomes}, {Gonz{\'a}lez Hern{\'a}ndez}, {Hughes},
  {Iwert}, {Kerber}, {Landoni}, {Lizon}, {Lovis}, {Maire}, {Mannetta},
  {Martins}, {Molaro}, {Monteiro}, {Moschetti}, {Oliveira}, {Zapatero Osorio},
  {Poretti}, {Rasilla}, {Santana Tschudi}, {Santos}, {Sosnowska}, {Sousa},
  {Tenegi}, {Toso}, {Vanzella}, \& {Viel}}]{espresso}
{M{\'e}gevand}, D., {Zerbi}, F.~M., {Di Marcantonio}, P., {et~al.} 2014, in
  \procspie, Vol. 9147, Ground-based and Airborne Instrumentation for Astronomy
  V, 91471H

\bibitem[{{Mortier} {et~al.}(2016){Mortier}, {Faria}, {Santos}, {Rajpaul},
  {Figueira}, {Boisse}, {Collier Cameron}, {Dumusque}, {Lo Curto}, {Lovis},
  {Mayor}, {Melo}, {Pepe}, {Queloz}, {Santerne}, {S{\'e}gransan}, {Sousa},
  {Sozzetti}, \& {Udry}}]{Mortier16}
{Mortier}, A., {Faria}, J.~P., {Santos}, N.~C., {et~al.} 2016, \aap, 585, A135

\bibitem[{{Radhakrishnan} {et~al.}(1984){Radhakrishnan}, {Abhyankar}, \&
  {Sarma}}]{1984BASI...12..182R}
{Radhakrishnan}, K.~R., {Abhyankar}, K.~D., \& {Sarma}, M.~B.~K. 1984, Bulletin
  of the Astronomical Society of India, 12, 182

\bibitem[{{Raskin} {et~al.}(2011){Raskin}, {van Winckel}, {Hensberge},
  {Jorissen}, {Lehmann}, {Waelkens}, {Avila}, {de Cuyper}, {Degroote},
  {Dubosson}, {Dumortier}, {Fr{\'e}mat}, {Laux}, {Michaud}, {Morren}, {Perez
  Padilla}, {Pessemier}, {Prins}, {Smolders}, {van Eck}, \&
  {Winkler}}]{Raskin2011}
{Raskin}, G., {van Winckel}, H., {Hensberge}, H., {et~al.} 2011, \aap, 526, A69

\bibitem[{{Ribas} {et~al.}(2002){Ribas}, {Arenou}, \&
  {Guinan}}]{2002AJ....123.2033R}
{Ribas}, I., {Arenou}, F., \& {Guinan}, E.~F. 2002, \aj, 123, 2033

\bibitem[{{Rodler} {et~al.}(2010){Rodler}, {K{\"u}rster}, \&
  {Henning}}]{Rodler10}
{Rodler}, F., {K{\"u}rster}, M., \& {Henning}, T. 2010, \aap, 514, A23

\bibitem[{{Rossiter}(1924)}]{Rossiter24}
{Rossiter}, R.~A. 1924, \apj, 60, doi:10.1086/142825

\bibitem[{{Sablowski} \& {Weber}(2017)}]{Sablowski17}
{Sablowski}, D.~P., \& {Weber}, M. 2017, \aap, 597, A125

\bibitem[{{Smette} {et~al.}(2015){Smette}, {Sana}, {Noll}, {Horst}, {Kausch},
  {Kimeswenger}, {Barden}, {Szyszka}, {Jones}, {Gallenne}, {Vinther},
  {Ballester}, \& {Taylor}}]{Smette2015}
{Smette}, A., {Sana}, H., {Noll}, S., {et~al.} 2015, \aap, 576, A77

\bibitem[{{Valenti} \& {Fischer}(2005)}]{Valenti05}
{Valenti}, J.~A., \& {Fischer}, D.~A. 2005, \apjs, 159, 141

\bibitem[{{Waldmann}(2012)}]{Waldmann12}
{Waldmann}, I.~P. 2012, \apj, 747, 12

\bibitem[{{Waldmann} {et~al.}(2013){Waldmann}, {Tinetti}, {Deroo}, {Hollis},
  {Yurchenko}, \& {Tennyson}}]{Waldmann13}
{Waldmann}, I.~P., {Tinetti}, G., {Deroo}, P., {et~al.} 2013, \apj, 766, 7

\bibitem[{{Winn}(2010)}]{Winn10}
{Winn}, J.~N. 2010, {Exoplanet Transits and Occultations} (University of
  Arizona Press), 55--77

\end{thebibliography}

\end{document}